\documentclass[seceqn,secthm]{elsart}
\usepackage{amssymb}\usepackage{amsmath}
\usepackage[all, knot]{xy}
\xyoption{arc} \xyoption{color}\xyoption{line}
\usepackage{mathrsfs}

\renewcommand{\leq}{\leqslant}
\renewcommand{\geq}{\geqslant}

\newcommand{\supp}{\operatorname{supp}}
\newcommand{\diag}{\operatorname{diag}}

\journal{arXiv:1003.2538}

\begin{document}

\begin{frontmatter}

\title{Aguilar-Balslev-Combes theorem for the Laplacian on a manifold with an axial analytic asymptotically cylindrical end\thanksref{AKA}}
\author{Victor Kalvin}
\thanks[AKA]{This work was funded by grant N108898 awarded by the Academy of Finland.}
\ead{vkalvin\,@\,gmail.com}
\date{}

\begin{abstract} We develop the complex scaling for a manifold with an asymptotically cylindrical end under an assumption on the analyticity of the metric with respect to the axial coordinate of the end. We allow for arbitrarily slow convergence of the metric to its limit at infinity, and prove a variant of the Aguilar-Balslev-Combes theorem for the Laplacian $\Delta$ on functions. In the case of a manifold with (noncompact) boundary it is either the Dirichlet or the Neumann Laplacian. We introduce resonances as the discrete non-real eigenvalues of non-selfadjoint operators, obtained as deformations of the Laplacian by means of the complex scaling. The resonances are identified with the poles of the resolvent matrix elements $((\Delta-\mu)^{-1}F, G)$ meromorphic continuation in $\mu$  across the essential spectrum of $\Delta$, where $F$ and $G$ are elements of an explicitly given set of analytic vectors. The Laplacian has no singular continuous spectrum, the eigenvalues can accumulate only at thresholds.
\end{abstract}

\end{frontmatter}


\section{Introduction}

We consider a manifold  $(\mathcal M,\mathsf g)$ with an
asymptotically cylindrical end. This means that $\mathcal M$ is a
smooth non-compact manifold of the form $\mathcal M_c\cup (\Bbb
R_+\times\Omega)$, where $\mathcal M_c$ is a compact manifold, and
$\Bbb R_+\times \Omega$ is the Cartesian product of the positive
semi-axis $\Bbb R_+$ and a compact manifold $\Omega$, see
Fig.~\ref{manif0} and Fig.~\ref{manif}. Furthermore, the  metric
$\mathsf g$ asymptotically approaches at infinity  the product
metric $dx\otimes dx+\mathfrak h$ on the semi-cylinder $\Bbb
R_+\times\Omega$, where $\mathfrak h$ is a metric on $\Omega$.

Usually, when studying the Laplacian on the manifold $(\mathcal
M,\mathsf g)$, one imposes more e.g.~\cite{chr
ST,Guillope,Melrose,MelroseScat} or less
e.g.~\cite{DES,DEM,Edward,FroHislop,Mueller,IKL} restrictive
assumptions on the rate of convergence of the metric $\mathsf g$ to
its limit $dx\otimes dx+\mathfrak h$ at infinity. Our goal is to
study the case of arbitrarily slow convergence. With this aim in
mind we invoke the complex scaling method. We assume that the metric
$\mathsf g$ has an analytic continuation to a conical neighborhood
of the axis $\Bbb R_+$ of the semi-cylinder $\Bbb R_+\times \Omega$,
and this continuation tends to a certain limit at infinity. If
$\mathsf g$ meets these assumptions, then   $(\mathcal M,\mathsf g)$
is said to be a manifold with an axial analytic asymptotically
cylindrical end; for precise definitions see  Section~\ref{s0}. We
study the Laplacian $\Delta$ on functions in three generic cases: 1.
$\Delta$ is the Laplacian on a manifold without boundary; 2.
$\Delta$ is the Dirichlet Laplacian on a manifold with noncompact
boundary; 3. $\Delta$ is the Neumann Laplacian on a manifold with
noncompact boundary. We exclude from consideration manifolds with
compact boundaries, as they can be treated similarly to the case 1.
 In this paper we develop a universal
approach to all three cases. However the most complicated case 3 is
considered as a principal one.

Despite there are several papers utilizing different approaches to
the complex scaling in geometric aspects
e.g.~\cite{DES,DEM,Kalvin,MV1,MV2,S,WZ},  the complex scaling has
not been used in this setting before. Our approach originates from
the one in~\cite{Hunziker}, where the Aguilar-Balslev-Combes theorem
is proved for the scattering problem of $n$ electrons in the field
of $N$ fixed nuclei (see also~\cite{Hislop Sigal}). In this paper we
characterize the spectrum and resonances of the Laplacian $\Delta$
on a manifold  with an axial analytic asymptotically cylindrical
end, establishing an analog of the  Aguilar-Balslev-Combes theorem.
In particular, we introduce resonances as the discrete non-real
eigenvalues of elliptic m-sectorial operators. These operators are
obtained as deformations of the Laplacian by means of the complex
scaling. The resonances are identified with the poles of the
resolvent matrix elements $((\Delta-\mu)^{-1}F, G)$ meromorphic
continuation in $\mu$  across the essential spectrum of $\Delta$,
where $F$ and $G$ are elements of an explicitly given sufficiently
large set of analytic vectors, and $(\cdot,\cdot)$ is the global
inner product on $(\mathcal M,\mathsf g)$. It turns out that the
Laplacian has no singular continuous spectrum, and its eigenvalues
can accumulate only at thresholds. In particular, this paper
generalizes our results~\cite{Kalvin} on the Dirichlet Laplacian in
a domain with an axial analytic asymptotically cylindrical end. Let
us note that the approach~\cite{Kalvin}, being  significantly based
on analysis of operators in a global system of Cartesian
coordinates, cannot be utilized here. First, because the needed
system of coordinates may not exist. Secondly, because in the case
of the Neumann Laplacian the complex scaling deforms not only the
Laplacian itself, but also the operator of boundary conditions.

In the proof of the Aguilar-Balslev-Combes theorem we distinguish
two substantial steps: (i) Proof of  the analyticity of the
resolvent of the deformed operator with respect to a scaling
parameter; (ii) Localization of the essential spectrum of the
deformed operator.

On the step (i) we employ the theory of analytic perturbations due
to Kato~\cite{Kato}. We arrange the complex scaling so that the
corresponding deformations of the Laplacian belong to the class of
m-sectorial operators and form an analytic family with respect to a
scaling parameter.  This is archived by taking a complex scaling,
deforming the operators only in a sufficiently small neighborhood of
infinity, and leads to the analyticity of the resolvent with respect
to a scaling parameter. An additional difficulty here is due to the
fact that the complex scaling deforms the domain of the Neumann
Laplacian.  However the domain of the corresponding quadratic form
remains unchanged. For this reason, and also because it is
convenient to work with m-sectorial operators in terms of their
quadratic forms, our methods are based on the analysis of analytic
families of quadratic forms. We introduce the quadratic forms in a
coordinate free way through non-Hermitian (sectorial) deformations
of the global inner product on $(\mathcal M,\mathsf g)$. These
deformations are obtained by the complex scaling of the metric.

On the step (ii), localizing the essential spectrum of the deformed
Laplacian, we consider the domain of the unbounded operator as a
Hilbert space and the corresponding bounded operator.  We rely on a
direct verification of the Fredholm property of the bounded operator
with spectral parameter.  The verification can be based either on
construction of parametrices, or, alternatively, on the approach due
to Peetre~\cite{Peetre}. This elegant approach allows to avoid a
tedious procedure of construction of parametrices by proving some
global coercive estimates, which is widely used in the theory of
elliptic boundary value
problems~\cite{KozlovMaz`ya,KozlovMazyaRossmann,Lions Magenes,MP2}.

First of all we  arrange the complex scaling so that the
corresponding deformations of the Laplacian remain in the class of
elliptic operators.  This is also achieved by taking a complex
scaling, deforming the operators only in a sufficiently small
neighborhood of infinity. A substantial step here is to demonstrate
that the deformed Neumann Laplacian satisfies the
Shapiro-Lopatinski\v{\i} condition on the boundary. Once the
ellipticity  is established, we obtain global coercive estimates by
methods of the theory of non-homogeneous elliptic boundary value
problems~\cite{KozlovMaz`ya,KozlovMazyaRossmann,Lions Magenes,MP2}.
This implies a condition on the spectral parameter, necessary and
sufficient for the Fredholm property of the operator, and  localizes
the essential spectrum. Let us stress that this approach  does not
require any assumptions on the rate of convergence of the metric
$\mathsf g$ at infinity.

Under our assumptions on the metric $\mathsf g$, accumulations of
isolated and embedded eigenvalues of the Laplacian may occur. The
Aguilar-Balslev-Combes theorem  implies that the non-threshold
eigenvalues of the  Laplacian are of finite multiplicity, and  can
accumulate only at the thresholds. In the companion
paper~\cite{KalvinIII} we refine these results by proving that a)
the non-threshold eigenfunctions of the Laplacian  are of
exponential decay at infinity, b) the eigenvalues are of finite
multiplicity and can accumulate at the thresholds only from below.
We believe that our methods can be extended to other non-compact
manifolds with a sufficiently explicit structure at infinity and to
a class of general elliptic operators of arbitrary order.

In the following two sections readers will find a precise
description of the geometric situation we deal with, the
Aguilar-Balslev-Combes theorem for the Laplacian, and a discussion
of our results.

 We complete this section with  the
 structure of the present paper. In Section~\ref{s0} we introduce manifolds with axial analytic asymptotically cylindrical ends. Then in Section~\ref{s2} we formulate and discuss our results. All subsequent sections are devoted to the proof.
Thus in Section~\ref{s3} we deform the Riemannian global inner
product by means of the complex scaling. In terms of this
deformation we define a sesquilinear quadratic form associated with
the Laplacian  deformed by means of the complex scaling. In
Section~\ref{s4} we study the quadratic form. As a result we obtain
an estimate on the spectrum of the deformed Laplacian and the
analyticity of its resolvent with respect to a scaling parameter. In
Section~\ref{s5} we localize the essential spectrum of the deformed
Laplacian. Finally, in Section~\ref{s6} we construct  the resolvent
matrix elements meromorphic continuation  and complete the proof of
our results.

\section{Manifolds with axial analytic asymptotically cylindrical ends}\label{s0}
\begin{figure}[h]
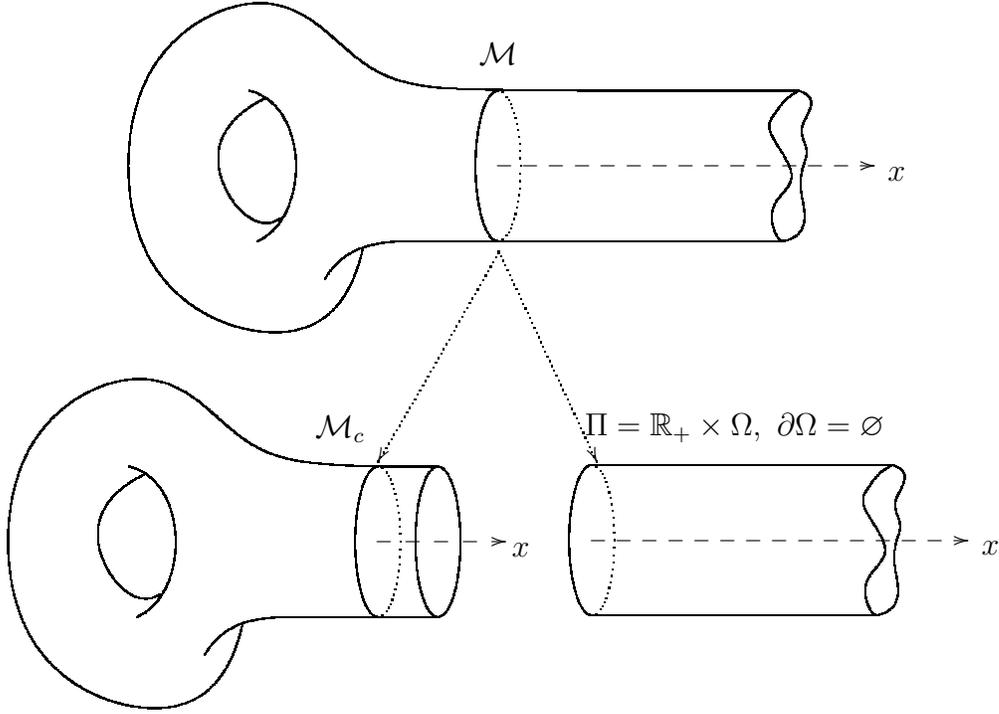

\[\xy (33,50)*{\xy
(40,10);(-18,-11)**\crv{(0,10)&(-13,10)&(-20,15)&(-30,25)&(-50,15)&(-50,-15)&(-30,-25)&(-20,-20)};
(38,-10);(-23,-15)**\crv{(0,-10)&(-10,-10)&(-20,-10)}; (-32,-10);
(-33,10)**\crv{(-25,-7)&(-25,8)}; (-29,-7);
(-31,9)**\crv{(-32,-9)&(-39,-2)&(-37,6)}; (40,10);
(38,-10)**\crv{(43,9)&(39,5)&(42,2)&(39,-2)&(42,-9)}; (40,10);
(38,-10)**\crv{(38,9)&(37,7)&(35,5)&(41,-2)&(35,-5)&(37,-10)};
{\ar@{-->} (0,0)*{}; (50,0)*{}}; (53,-1)*{x};
(0,0)*\ellipse(3,10){.}; (0,0)*\ellipse(3,10)__,=:a(-180){-};
(0,15)*{\mathcal M};
\endxy};
(0,0)*{\xy (8,10);
(-18,-11)**\crv{(0,10)&(-13,10)&(-20,15)&(-30,25)&(-50,15)&(-50,-15)&(-30,-25)&(-20,-20)};
(8,-10); (-23,-15)**\crv{(0,-10)&(-10,-10)&(-20,-10)}; (-32,-10);
(-33,10)**\crv{(-25,-7)&(-25,8)}; (-29,-7);
(-31,9)**\crv{(-32,-9)&(-39,-2)&(-37,6)}; (-5,15)*{\mathcal M_c};
{\ar@{-->} (0,0)*{}; (17,0)*{}}; (19,-1)*{x};
(0,0)*\ellipse(3,10){.}; (0,0)*\ellipse(3,10)__,=:a(-180){-};
(4,0)*\ellipse(3,10){-}; (4,0)*\ellipse(3,10)__,=:a(-180){-};
\endxy};
(70,4)*{\xy (0,10); (40,10)**\dir{-};(0,-10); (38,-10)**\dir{-};
(0,0)*\ellipse(3,10){.}; (0,0)*\ellipse(3,10)__,=:a(-180){-};
(19,15)*{\Pi=\Bbb R_+\times\Omega,\
\partial\Omega=\varnothing};(0,15)*{\vphantom{\mathcal M_c}};
(40,10); (38,-10)**\crv{(43,9)&(39,5)&(42,2)&(39,-2)&(42,-9)};
(40,10);
(38,-10)**\crv{(38,9)&(37,7)&(35,5)&(41,-2)&(35,-5)&(37,-10)};
{\ar@{-->} (0,0)*{}; (50,0)*{}}; (53,-1)*{x};
(0,0)*\ellipse(3,10){.}; (0,0)*\ellipse(3,10)__,=:a(-180){-};
\endxy};
{\ar@{.>} (31,39)*{};(15,11)*{}};{\ar@{.>} (31,39)*{};(44,11)*{}};
\endxy\]
\caption{Representation $\mathcal M=\mathcal M_{c}\cup\Pi$ of a
manifold $\mathcal M$ without boundary.}\label{manif0}
\end{figure}

Let $\Omega$ be a smooth compact  $n$-dimensional manifold with
smooth boundary $\partial\Omega$ or without it. Denote by $\Pi$ the
semi-cylinder $\Bbb R_+\times \Omega$, where $\Bbb R_+$ is the
positive  semi-axis, and $\times$ stands for the Cartesian product.
Consider  a  smooth oriented connected $n+1$-dimensional  manifold
$\mathcal M$   representable in the form $\mathcal M=\mathcal
M_c\cup \Pi$, where $\mathcal M_c$ is a smooth compact manifold with
boundary, cf. Fig.~\ref{manif0} and Fig.~\ref{manif}.  It will be
convenient to assume that $(0,1)\times\Omega\subset \mathcal
M_c\cap\Pi$. We exclude from consideration the case of a manifold
$\mathcal M$ with compact boundary $\partial\mathcal M$, assuming
that $\partial\mathcal M=\varnothing$, if
$\partial\Omega=\varnothing$.
\begin{figure}[h]
\[\xy (33,50)*{\xy
(40,10);(-18,-11)**\crv{(0,10)&(-13,10)&(-20,15)&(-30,25)&(-50,15)&(-50,-15)&(-30,-25)&(-20,-20)};
(38,-10);(-23,-15)**\crv{(0,-10)&(-10,-10)&(-20,-10)}; (-32,-10);
(-33,10)**\crv{(-25,-7)&(-25,8)}; (-29,-7);
(-31,9)**\crv{(-32,-9)&(-39,-2)&(-37,6)}; (40,10);
(38,-10)**\crv{(43,9)&(39,5)&(42,2)&(39,-2)&(42,-9)}; (40,10);
(38,-10)**\crv{(38,9)&(37,7)&(35,5)&(41,-2)&(35,-5)&(37,-10)};
{\ar@{-->} (0,0)*{}; (50,0)*{}}; (53,-1)*{x};
(0,0)*\ellipse(3,10){.}; (0,0)*\ellipse(3,10)__,=:a(-180){-};
(0,15)*{\mathcal M};
\endxy};
(0,0)*{\xy (8,10);
(-18,-11)**\crv{(0,10)&(-13,10)&(-20,15)&(-30,25)&(-50,15)&(-50,-15)&(-30,-25)&(-20,-20)};
(8,-10); (-23,-15)**\crv{(0,-10)&(-10,-10)&(-20,-10)}; (-32,-10);
(-33,10)**\crv{(-25,-7)&(-25,8)}; (-29,-7);
(-31,9)**\crv{(-32,-9)&(-39,-2)&(-37,6)}; (-5,15)*{\mathcal M_c};
(10,-7); (10,6)**\crv{(10,-6)&(14,-2)&(15,1)&(10,4)&(10,7)}; (8,10);
(13,2)**\crv{(12,10)}; (8,-10); (13,-3)**\crv{(12,-10)}; {\ar@{-->}
(0,0)*{}; (17,0)*{}}; (19,-1)*{x}; (0,0)*\ellipse(3,10){.};
(0,0)*\ellipse(3,10)__,=:a(-180){-};
\endxy};
(70,4)*{\xy (0,10); (40,10)**\dir{-};(0,-10); (38,-10)**\dir{-};
(0,0)*\ellipse(3,10){.}; (0,0)*\ellipse(3,10)__,=:a(-180){-};
(21,15)*{\Pi=\Bbb R_+\times\Omega,\
\partial\Omega\neq\varnothing};(0,15)*{\vphantom{\mathcal M_c}};
(40,10); (38,-10)**\crv{(43,9)&(39,5)&(42,2)&(39,-2)&(42,-9)};
(40,10);
(38,-10)**\crv{(38,9)&(37,7)&(35,5)&(41,-2)&(35,-5)&(37,-10)};
{\ar@{-->} (0,0)*{}; (50,0)*{}}; (53,-1)*{x};
(0,0)*\ellipse(3,10){.}; (0,0)*\ellipse(3,10)__,=:a(-180){-};
\endxy};
{\ar@{.>} (31,39)*{};(15,11)*{}};{\ar@{.>} (31,39)*{};(44,11)*{}};
\endxy\]
\caption{Representation $\mathcal M=\mathcal M_{c}\cup\Pi$ of a
manifold $\mathcal M$ with boundary.}\label{manif}
\end{figure}

Let $\mathsf g\in C^\infty \mathrm T^*\mathcal M^{\otimes 2}$ be a
Riemannian metric on $\mathcal M$. We identify the cotangent bundle
$\mathrm T^*\Pi$ with the tensor product $\mathrm T^*\Bbb
R_+\otimes\mathrm T^*\Omega$   via the natural isomorphism induced
by the product structure on $\Pi$. This together with the
trivialization $\mathrm T^*\Bbb R_+=\{(x,a\, dx): x\in\Bbb R_+,a\in
\Bbb R\}$ implies that any metric   $\mathsf g$ can be represented
on $\Pi$ in the form
\begin{equation}\label{split}
\mathsf g\!\upharpoonright_{\Pi}=\mathfrak g_0 dx\otimes dx+2
\mathfrak g_1\otimes dx +\mathfrak g_2, \quad\mathfrak g_k(x)\in
C^\infty\mathrm T^*\Omega^{\otimes k}.
\end{equation}

Denote by  $\Bbb C\mathrm T^*\Omega^{\otimes k}$  the tensor power
of the complexified cotangent bundle $\Bbb C\mathrm T^*\Omega$ with
the fibers $\Bbb C\mathrm T_{\mathrm y}^*\Omega=\mathrm T_{\mathrm
y}^*\Omega\otimes\Bbb C$. In what follows $C^m$ stands for sections
of complexified bundles, e.g. we write $C^\infty \mathrm
T^*\Omega^{\otimes k}$ and $C^1 \mathrm T^*\Omega^{\otimes k}$
instead of $C^\infty\Bbb C \mathrm T^*\Omega^{\otimes k}$ and
$C^1\Bbb C\mathrm T^*\Omega^{\otimes k}$. We equip the space
$C^1\mathrm T^*\Omega^{\otimes k}$ with the norm
\begin{equation}\label{Enorm}
\|\cdot\|_{\mathfrak e}=\max_{\mathrm
y\in\Omega}\bigl(|\cdot|_{\mathfrak e}(\mathrm
y)+|D\cdot|_{\mathfrak e}(\mathrm y)\bigr),
\end{equation}
where $\mathfrak e$ is a Riemannian metric on $\Omega$,
$|\cdot|_{\mathfrak e}(\mathrm y)$ is the norm induced by
$\mathfrak e$ in the fiber $\Bbb C\mathrm T_{\mathrm
y}^*\Omega^{\otimes k}$, and $ D: C^1\mathrm T^*\Omega^{\otimes
k}\to C^0\mathrm T^*\Omega^{\otimes k+1}$ is the Levi-Civita
connection on the manifold $(\Omega,\mathfrak e)$.

\begin{defn}\label{ACE} We say that $(\Pi,\mathsf g\!\upharpoonright_{\Pi})$ is an axial analytic asymptotically cylindrical end, if the following conditions hold:
\begin{itemize}
\item[i.]  The functions   $ x\mapsto \mathfrak g_k(x)\in C^\infty\mathrm T^*\Omega^{\otimes k}$ in~\eqref{split}   extend  by analyticity in $x$ from $\Bbb R_+$ to the sector $\Bbb S_\alpha=\{z\in\Bbb C:|\arg z|<\alpha\}$ with some  $\alpha>0$.
\item[ii.] As $z$ tends to infinity in $\Bbb S_\alpha$ the function $\mathfrak g_0(z)$ uniformly converges to $1$ in the norm of $C^1(\Omega)$, the tensor field $\mathfrak g_1(z)$ uniformly converges to zero in the norm of $C^1\mathrm T^*\Omega$, and the tensor field $\mathfrak g_2(z)$ uniformly converges to a Riemannian metric $\mathfrak h$ on $\Omega$ in the norm of $C^1\mathrm T^*\Omega^{\otimes 2}$.
\end{itemize}
\end{defn}

Definition~\ref{ACE} is independent of the metric $\mathfrak e$
defining the norm in $C^1\mathrm T^*\Omega^{\otimes k}$.

In this paper we consider a manifold $(\mathcal M,\mathsf g)$  with
an axial analytic asymptotically cylindrical end  $(\Pi,\mathsf
g\!\upharpoonright_\Pi)$. Definition~\ref{ACE} allows for
arbitrarily slow convergence  of the metric $\mathsf g$  to the
product metric $\overset{\infty}{\mathsf g}=dx\otimes dx+\mathfrak
h$ at infinity.

We will often work in local coordinates on $\Omega$. By $\{\mathscr
U_j, \kappa_j\}$ we denote a finite atlas on $\Omega$. Let $y\in\Bbb
R^n$ be a  system of local coordinates in a neighborhood $\mathscr
U_j$. In the case $\partial\Omega\cap\mathscr U_j\neq\varnothing$ we
suppose that all $y$ in  $\kappa_j[\partial\Omega\cap\mathscr U_j]$
(i.e. in the image of the set $\partial\Omega\cap\mathscr U_j$ under
the diffeomorphism $\kappa_j$) are of the form  $y=(y',y_n)$ with
$y'\in \Bbb R^{n-1}$ and $y_n\geq 0$; moreover,  the set $\mathscr
U_j\cap\partial\Omega$ is defined by the equality $y_n=0$.
 We will use the notations  $\partial_{y_m}=\frac \partial {\partial y_{m}}$  and
   $\partial^r_{y}=\partial^{r_1}_{y_1}\partial^{r_2}_{y_2}\dots\partial^{r_n}_ {y_n}$,
   where $r=(r_1,\dots,r_n)$ is a multiindex, and $|r|=\sum r_m$. Below we give a definition of an axial analytic asymptotically  cylindrical end  in terms of local coordinates.
\begin{defn}\label{R1}
Let $\mathbf g(x,y)$ be the matrix corresponding to the
representation of the metric $\mathsf g$ in the coordinates
$(x,y)\in \Bbb R_+\times \kappa_j[\Omega\cap\mathscr U_j]$ on $\Pi$.
Then $(\Pi,\mathsf g\!\upharpoonright_{\Pi})$ is an  axial analytic
asymptotically cylindrical end, if in every neighborhood $\mathscr
U_j$ the matrix elements
$$
\Bbb R_+\ni x\mapsto \mathbf g_{\ell m}(x,\cdot)\in
C^\infty(\kappa_j[\Omega\cap\mathscr U_j])
$$
have analytic continuations from $\Bbb R_+$ to the sector $\Bbb
S_\alpha$, and the stabilization condition
 \begin{equation}\label{stab}
   \sum_{|r|\leq 1}\bigl\|\partial_y^r\bigl(\mathbf g(z,y)-\overset{\infty}{\mathbf g}(y)\bigr)\bigr\|_{2}\to 0\text{ as } |z|\to\infty
   \end{equation}
holds uniformly in $z\in\mathbb S_\alpha$ and
$y\in\kappa_j[\Omega\cap\mathscr U_j]$. Here  $\|\cdot\|_{2}$ is
the  matrix norm  $\|\mathbf g\|_{2}=\sqrt{\sum_{\ell,m=0}^n
|\mathbf g_{\ell m}|^2}$, and the matrix $\overset{\infty}{\mathbf
g}(y)$ corresponds to the representation of the product metric
$\overset{\infty}{\mathsf g}=dx\otimes dx+\mathfrak h$ on $\Pi$ in
the  coordinates $(x,y)$.
\end{defn}
The proof of equivalence of Definitions~\ref{ACE} and~\ref{R1} is
postponed to Section~\ref{s3}.

\begin{figure}[h]
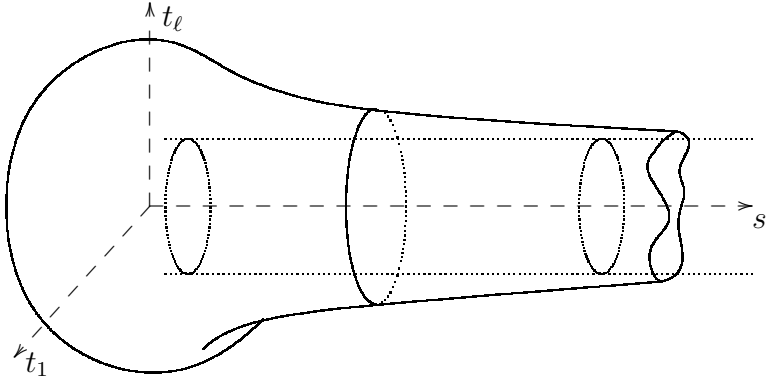

\[\xy  (40,10);(-15,-15)**\crv{(0,12)&(-13,15)&(-20,18)&(-30,25)&(-50,15)&(-50,-15)&(-30,-25)&(-20,-20)};
(38,-10);(-23,-19)**\crv{(0,-13)&(-10,-14)&(-20,-16)};
(40,10); (38,-10)**\crv{(43,9)&(39,5)&(42,2)&(39,-2)&(42,-9)};
(40,10);
(38,-10)**\crv{(38,9)&(37,7)&(35,5)&(41,-2)&(35,-5)&(37,-10)};
{\ar@{-->} (-30,0)*{}; (50,0)*{}}; (51,-2)*{s};
(0,0)*\ellipse(4,13){.}; (0,0)*\ellipse(4,13)__,=:a(-180){-};
{\ar@{-->} (-30,0)*{}; (-30,27)*{}}; (-27,25)*{t_\ell}; {\ar@{-->}
(-30,0)*{}; (-48,-20)*{}}; (-45,-21)*{t_1};
{\ar@{.}(-28,9); (50,9)}; {\ar@{.}(-28,-9); (50,-9)};
(-25,0)*\xycircle(3,9){.}; (30,0)*\xycircle(3,9){.};
\endxy\]
\caption{Domain $\mathcal G\subset \Bbb R^{n+1}$ with an
asymptotically cylindrical end.}\label{domain}
\end{figure}

Let us give some illustrative examples of  manifolds  $(\mathcal
M,\mathsf g)$ with axial analytic asymptotically cylindrical ends.
Let $ \Omega$ be a bounded domain in $\Bbb R^n$ with smooth boundary
($\Omega$ is a bounded interval  in the case $n=1$). By  $(x,y)$ and
$(s,t)$ we denote the Cartesian coordinates in $\Bbb R^{n+1}$, such
that $x,s\in\Bbb R$ and $y,t\in\Bbb R^n$.  Consider a closed domain
$\mathcal M$ with  smooth boundary, such that the set  $\{(x,y)\in
{\mathcal M}: x\leq 0\}$ is a bounded subset of the half-space
$\{(x,y)\in \Bbb R^{n+1}: x>-2\}$, and the set $\{(x,y)\in {\mathcal
M}:x> 0\}$ is the semi-cylinder $\Pi=\{(x,y): x\in\Bbb R_+,
y\in\Omega\}$. Let the domain
$$
\mathcal G=\{(s,t)\in\Bbb R^{n+1}: (s,t)=\phi(x,y), (x,y)\in\mathcal
M\}
$$
be the image of $\mathcal M$ under a diffeomorphism $\phi$. Assume
that $\phi$ satisfies the following conditions:
\begin{enumerate}
\item[\it i.] the
function $x\mapsto \phi(x,\cdot)\in C^\infty(\Omega, \Bbb C^{n+1})$
has an analytic continuation from $\mathbb R_+$ to the sector $\Bbb
S_\alpha$ with some $\alpha>0$;
 \item[\it ii.] the elements $\phi'_{\ell m}(z,\cdot)$ of the Jacobian matrix $\phi'=\{\phi'_{\ell m}\}_{\ell,m=0}^{n}$ uniformly tend to the
Kronecker delta $\delta_{\ell m}$ in the space $C^\infty(\Omega)$
as $z$ tends to infinity in the sector $\Bbb S_\alpha$.
\end{enumerate}
Here $C^\infty(\Omega, \Bbb C^{n+1})$ stands for the space of smooth
functions acting from $\Omega$ to $\Bbb C^{n+1}$. Let $\mathsf
g=\phi^*\overset{\infty}{\mathsf g}$ be the pullback of the
Euclidean metric $\overset{\infty}{\mathsf g}$ on $\mathcal G$ by
the diffeomorphism $\phi$. Then the metric
$$\mathsf g=(\phi'_{00})^2dx \otimes dx+2\sum_{ m=1}^n\sum_{\ell=0}^n \phi'_{0 \ell }\phi'_{\ell m}\, dx\otimes d y_m+\sum_{m,j=1}^n\sum_{\ell=0}^n\phi'_{m \ell }\phi'_{\ell j} \,dy_m\otimes d y_j$$
on $\mathcal M$ has all properties required in Definitions~\ref{ACE}
and~\ref{R1}.
 For instance, we can take
 \begin{equation}\label{dif1}
\phi(x,y)=\left( x,(x+3)^\beta a+ (1+(x+3)^{\gamma} )y\right),\quad
a\in\Bbb R^n,\ \beta<1,\ \gamma<0,
\end{equation}
\begin{equation}\label{dif2}
\phi(x,y)=\left(\int_0^x \bigl(1+1/\log(\tilde x+4)\bigr)\, d\tilde
x, \bigl(1+1/\log(x+5)\bigr)y\right).
\end{equation}
In the case~\eqref{dif1}  the boundary $\partial\mathcal G$
asymptotically approaches at infinity the bent semi-cylinder
$\{(s,t): s\in\Bbb R_+, t-(s+2)^\beta a\in\Omega\}$, and in the
case~\eqref{dif2} the boundary $\partial\mathcal G$ asymptotically
approaches at infinity the semi-cylinder $\Bbb
R_+\times\partial\Omega$, cf. Fig.~\ref{domain}.
Evidently, the surface $\partial\mathcal G$ in  $\Bbb R^{n+1}$,
$n\geq 2$, can be viewed as  a manifold $(\partial \mathcal
M,\mathsf g\!\upharpoonright_{\partial \mathcal M})$ without
boundary. Then   $(\Bbb R_+\times\partial\Omega,\mathsf
g\!\upharpoonright_{\Bbb R_+\times\partial\Omega})$ is an axial
analytic asymptotically cylindrical end. Let us remark here that to
the best knowledge of the author the manifold $(\mathcal M,\mathsf
g)$, where $\mathsf g$ is the pullback of the Euclidean metric
$\overset{\infty}{\mathsf g}$ by the diffeomorphism~\eqref{dif2}, is
not covered by any of the previously known results on spectral
properties of the Laplacian due to very slow convergence of
$\mathsf g$ to $\overset{\infty}{\mathsf g}$ at infinity.

\section{Aguilar-Balslev-Combes theorem}\label{s2}
In this section we  formulate and discuss the Aguilar-Balslev-Combes
theorem.  As preliminaries to the theorem we introduce a deformation
of the Laplacian by means of the complex scaling, define the
thresholds, and introduce a sufficiently large set of analytic
vectors.

Let $(\mathcal M,\mathsf g)$ be  a  manifold
 with an axial analytic asymptotically
cylindrical end $(\Pi,\mathsf g\!\upharpoonright_{\Pi})$. In this
paper we use the complex scaling $x\mapsto x+\lambda \mathsf s_R(x)$
along the axis $\Bbb R_+$ of the semi-cylinder $\Pi=\Bbb
R_+\times\Omega$. Here $\lambda$ is a complex scaling parameter, and
$\mathsf s_R(x)=\mathsf s(x-R)$ is a scaling function with  a
sufficiently large number $R>0$ and a smooth function $\mathsf s$
possessing the properties:
\begin{equation}\label{ab}
\begin{aligned}
&\mathsf s(x)=0 \text{ for all } x\leq 1,\\
&0\leq \mathsf s'(x)\leq 1 \text{ for all } x\in\Bbb R, \text{ and }
\mathsf s'(x)= 1 \text{ for large
 } x,
\end{aligned}
\end{equation}
where $\mathsf s'=\partial \mathsf s/\partial x$.  The function
$\Bbb R_+\ni x\mapsto x+\lambda \mathsf s_R(x)$ is invertible for
all real $\lambda\in(-1,1)$, and thus defines the selfdiffeomorphism
$$
\Pi\ni (x,\mathrm y)\mapsto\varkappa_\lambda (x,\mathrm
y)=(x+\lambda \mathsf s_R(x),\mathrm y)\in \Pi
$$
of the  semi-cylinder $\Pi$. This selfdiffeomorphism scales the
semi-cylinder along its axis. We extend $\varkappa_\lambda$ to a
selfdiffeomorphism of the manifold $\mathcal M$ by setting
$\varkappa_\lambda (p):=p$ for all points $p\in\mathcal
M\setminus\Pi$. As a result we get Riemannian manifolds $(\mathcal
M, \mathsf g_\lambda)$ parametrized by
 $\lambda\in(-1,1)$, where the metric
$\mathsf g_\lambda=\varkappa_\lambda^*\mathsf g$ is the pullback of
the metric $\mathsf g$ by  $\varkappa_\lambda$. In the case
$\lambda=0$ the scaling is not applied and $\mathsf g\equiv\mathsf
g_0$. Let us remark that  $\varkappa_\lambda$ and  $\mathsf
g_\lambda$ both depend on the parameter $R$, however we do not
indicate this for brevity of notations.

Let ${^\lambda\!\Delta}$ be the Laplacian on functions associated
with the metric $\mathsf g_\lambda$ on $\mathcal M$. In the case of
a manifold $\mathcal M$ with (noncompact) boundary we also consider
the directional derivative  $^\lambda\!\partial_\nu $ on
$\partial\mathcal M$ taken along the unit inward normal vector given
by the metric $\mathsf g_\lambda$.  It turns out that the Laplacian
${^\lambda\!\Delta}:C_c^\infty(\mathcal M)\to C_c^\infty(\mathcal
M)$ and the operator $^\lambda\!\partial_\nu:C_c^\infty(\mathcal
M)\to C_c^\infty(\partial\mathcal M)$ of the Neumann boundary
condition extend by analyticity in $\lambda$  to  the disk
\begin{equation}\label{disk}
\mathcal D_\alpha=\{\lambda\in\mathbb C: |\lambda|<\sin\alpha<
1/\sqrt{2}\},
\end{equation}
where $\alpha<\pi/4$ is some angle for which the conditions of
Definition~\ref{ACE} hold.

Introduce the Hilbert space $L^2(\mathcal M)$ as the completion of
the set $C_c^\infty(\mathcal M)$ with respect to the norm
$\|\cdot\|=\sqrt{(\cdot,\cdot)}$, where $(\cdot,\cdot)$ is the
global inner product induced on $\mathcal M$ by the metric $\mathsf
g$. From now on we  consider ${^\lambda\!\Delta}$ with
$\lambda\in\mathcal D_\alpha$ as an operator in $L^2(\mathcal M)$,
initially defined on a dense in $L^2(\mathcal M)$ core $\mathbf
C({^\lambda\!\Delta})$. The operator ${^\lambda\!\Delta}$ is a
deformation of the  Laplacian $\Delta\equiv {^0\!\Delta}$ by means
of the complex scaling.
\begin{defn}\label{d1} In the case
$\partial\mathcal M=\varnothing$  the core  $\mathbf
C({^\lambda\!\Delta})$ coincides with the set $C^\infty_c(\mathcal
M)$ of all smooth compactly supported functions on $\mathcal M$. In
the case of the Neumann (resp. Dirichlet) Laplacian the core
$\mathbf C({^\lambda\!\Delta})$ consists of the functions $u \in
C^\infty_c({\mathcal M})$ satisfying the deformed Neumann boundary
condition ${^\lambda\!\partial_\nu} u=0$ (resp. the Dirichlet
boundary condition $u\!\upharpoonright_{\partial \mathcal M}=0$).
\end{defn}
In general, the operator ${^\lambda\!\partial_\nu}$ and therefore
the core $\mathbf C({^\lambda\!\Delta})$ depend on the scaling
parameter $\lambda$ (it is not the case, if $(\mathcal M,\mathsf g)$
is a manifold with a cylindrical end).

Let $(\Omega,\mathfrak h)$ be the same compact Riemannian manifold
as in Definition~\ref{ACE}. Recall that we exclude from
consideration the case of a manifold $\mathcal M$ with compact
boundary $\partial\mathcal M$, i.e. the equalities $\partial\mathcal
M=\varnothing$ and $\partial\Omega=\varnothing$ hold only
simultaneously.  If $\partial \mathcal M\neq \varnothing$ and
$\Delta$ is the Dirichlet (resp. Neumann) Laplacian on $(\mathcal
M,\mathsf g)$, then  by $\Delta_\Omega$ we denote the Dirichlet
(resp. Neumann)  Laplacian on  $(\Omega,\mathfrak h)$. If
$\partial\mathcal M=\varnothing$, then $\Delta_\Omega$ is the
Laplacian on the manifold $(\Omega,\mathfrak h)$ without boundary.
Let $L^2(\Omega)$ be the Hilbert space of all square summable
functions on $(\Omega,\mathfrak h)$. As is well-known, the spectrum
of the operator $\Delta_\Omega$ in $L^2(\Omega)$ consists of
infinitely many  nonnegative isolated eigenvalues. Let
$\nu_1<\nu_2<\dots$ be the distinct eigenvalues of $\Delta_\Omega$.
By definition
$\{\nu_j\}_{j=1}^\infty$ is the set of thresholds of $\Delta$. 

 Before formulating our results, we introduce analytic vectors.  Consider the algebra  $\mathscr E$ of all entire functions $\mathbb C\ni z\mapsto f(z)\in C^\infty(\Omega)$ with the following property:
in any sector $|\Im z|\leq (1-\epsilon) \Re z$ with $\epsilon>0$
the value $\|f(z);{L^2(\Omega)}\|$ decays faster than any  inverse
power of $\Re z$  as $\Re z\to+\infty$.  Examples of  functions
$f\in \mathscr E$ are $f(z)=e^{-\gamma z^2}P(z)$, where  $\gamma>0$
and $P(z)$ is an arbitrary polynomial  in $z$ with  coefficients in
$C^\infty(\Omega)$. We say that $F\in L^2(\mathcal M)$ is an
analytic vector, if $F=f$ on $\Pi$  for some $f\in\mathscr E$. The
set of all analytic vectors is denoted by $\mathcal A$. Later on we
will show that the set $\mathcal A$ is  dense in  $L^2(\mathcal M)$.

\begin{thm}\label{T1} Consider a manifold $(\mathcal M,\mathsf g)$ with an axial analytic asymptotically cylindrical end $(\Pi,\mathsf g\!\upharpoonright_{\Pi})$. Assume that the scaling parameter $\lambda$ is in the disk~\eqref{disk}, where $\alpha<\pi/4$ is  the same as in Definition~\ref{ACE}. Let the deformation  $^\lambda\!\Delta$ of the Laplacian $\Delta$ on $(\mathcal M,\mathsf g)$ be constructed with a smooth scaling function $\mathsf s_R(x)=\mathsf s(x-R)$, where $\mathsf s\in C^\infty(\Bbb R)$ meets the conditions~\eqref{ab}, and $R>0$ is sufficiently large. Then the following assertions are valid.

\begin{itemize}
\item[1.] The operator ${^\lambda\!\Delta}$ in  $L^2(\mathcal M)$, initially defined on the core $\mathbf C({^\lambda\!\Delta})$, admits a closure, denoted by the same symbol ${^\lambda\!\Delta}$. The unbounded operator ${^\lambda\!\Delta}$ with $\lambda\neq 0$ is non-selfadjoint, however the Laplacian $\Delta\equiv{^0\!\Delta}$  is selfadjoint.
\item[2.] The spectrum $\sigma(^\lambda\!\Delta)$ of the operator  $^\lambda\!\Delta$  is independent of the choice of  $\mathsf s(x)$.

\item[3.]  $\mu$ is a point of the essential spectrum $\sigma_{ess}({^\lambda\!\Delta})$ of ${^\lambda\!\Delta}$, if and only if
\begin{equation}\label{eq9}
\mu= \nu_j\text{ or }\arg(\mu-\nu_j)=-2\arg (1+\lambda)\text{ for
some }j\in\mathbb N,
\end{equation}
where $\{\nu_j\}_{j=1}^\infty$ is the set of thresholds of $\Delta$.

\item[4.]  $\sigma({^\lambda\!\Delta})=\sigma_{ess}({^\lambda\!\Delta})\cup\sigma_d({^\lambda\!\Delta})$, where $\sigma_d(^\lambda\!\Delta)$ is the discrete spectrum of ${^\lambda\!\Delta}$.

\item[5.] Let $\mu\in\sigma_d({^\lambda\!\Delta})$. As $\lambda$ changes continuously in the disk $\mathcal D_\alpha$, the point $\mu$ remains in $\sigma_d({^\lambda\!\Delta})$ as long as $\mu\in\mathbb C\setminus\sigma_{ess}({^\lambda\!\Delta})$.

\item[6.] Let $(\cdot,\cdot)$ stand for the inner product in $L^2(\mathcal M)$. Then for any $F,G\in \mathcal A$ the analytic  function
$\mathbb C\setminus\overline{\mathbb R_+}\ni\mu\mapsto\bigl(
(\Delta-\mu)^{-1}F,G\bigr)$ has a meromorphic continuation to the
set $\mathbb C\setminus\sigma_{ess}({^\lambda\!\Delta})$. Moreover,
$\mu$ is a pole of the meromorphic continuation with some  $F,G\in
\mathcal A$, if and only if $\mu\in \sigma_d({^\lambda\!\Delta})$.

\item[7.] A point  $\mu\in \mathbb R$, such that $\mu\neq\nu_j$ for all $j\in\mathbb N$, is an eigenvalue of the Laplacian $\Delta$, if and only if $\mu\in \sigma_d({^\lambda\!\Delta})$ with $\Im\lambda\neq 0$.

\item[8.] The  Laplacian $\Delta$  has no singular
continuous spectrum.
\end{itemize}
\end{thm}

A similar result for the stationary Schr\"{o}dinger operator in
$\mathbb R^n$ is known as the Aguilar-Balslev-Combes theorem, see
e.g.~\cite{Hunziker,Hislop Sigal,Simon Reed iv} and references
therein.

\begin{figure}[h]
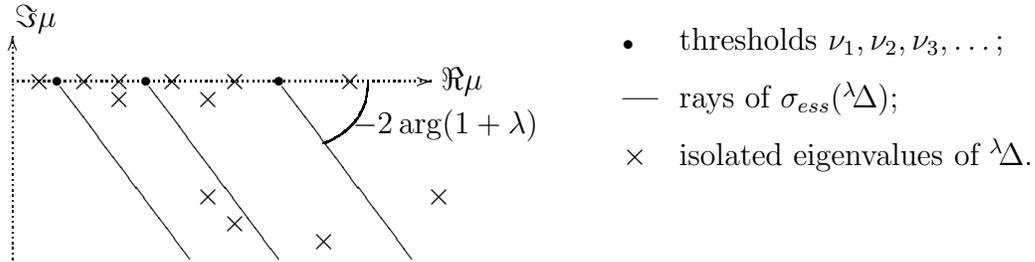

\[
\xy (0,0)*{\xy0;/r.28pc/:{\ar@{.>}(0,10);(50,10)*{\ \Re \mu}};
{\ar@{.>}(0,-10);(0,15)}; (2,17)*{\ \Im \mu};
(3,10)*{\times};(8,10)*{\times};(12,10)*{\times};(18,10)*{\times};(25,10)*{\times};(38,10)*{\times};{\ar@{-}(5,10)*{\scriptstyle\bullet};(20,-10)};{\ar@{-}(15,10)*{\scriptstyle\bullet};(30,-10)};
{(30,10)*{\scriptstyle\bullet};(45,-10)*{} **\dir{-}};
(40,10);(35,3)**\crv{(40,5)*{\quad\quad\quad\quad\quad
-2\arg(1+\lambda)}};
(35,-8)*{\times};(25,-6)*{\times};(22,-3)*{\times};(22,8)*{\times};(12,8)*{\times};(48,-3)*{\times}\endxy};
(74,0)*{\xy (20,20)*{
\begin{array}{ll}
     {\scriptstyle\bullet } & \text{ thresholds $\nu_1,\nu_2,\nu_3,\dots$;} \\
    \text{\bf---} & \text{ rays of $\sigma_{ess}({^\lambda\!\Delta})$;} \\
    {\times} &  \text{ isolated eigenvalues of $^\lambda\!\Delta$.}\\
\end{array}};
\endxy};
\endxy
\]
\caption{Spectral portrait of the deformed Dirichlet Laplacian
$^\lambda\!\Delta$, $\Im\lambda>0$.}\label{fig5}
\end{figure}

 The spectral portrait of the operator
${^\lambda \!\Delta}$ is depicted on Fig.~\ref{fig5}.
 As the parameter
$\lambda$ varies, the ray $\arg(\mu-\nu_j)=-2\arg (1+\lambda)$ of
the essential spectrum
 $\sigma_{ess}({^\lambda\!\Delta})$ rotates about the threshold $\nu_j$,
and sweeps the sector $|\arg (\mu-\nu_j)|<2\alpha$. By the
assertion~{\it 4} the eigenvalues of ${^\lambda \!\Delta}$ outside
of the sector $|\arg (\mu-\nu_1)|<2\alpha$ do not change, hence they
are the discrete eigenvalues of the selfadjoint Laplacian $\Delta$.
All other discrete eigenvalues of ${^\lambda \!\Delta}$ belong to
the sector $|\arg (\mu-\nu_1)|<2\alpha$. As $\lambda$ varies, they
remain unchanged until they are covered by one of the rotating rays
of the essential spectrum. Conversely, new eigenvalues can be
uncovered by the rotating rays. In the case $\Im\lambda\geq 0$
(resp. $\Im\lambda\leq 0$) the operator ${^\lambda \!\Delta}$
 cannot have eigenvalues in the half-plane $\Im
\mu>0$ (resp. $\Im \mu<0$). Indeed, by the assertion~{\it 4} a
number $\mu$ with $\Im \mu>0$   is an eigenvalue of ${^\lambda
\!\Delta}$ with $\Im\lambda\geq 0$, if and only if $\mu$ is an
eigenvalue of $\Delta$, but  the Laplacian $\Delta$ cannot have
non-real eigenvalues as a selfadjoint operator. Further, by the
assertion~{\it 6} the real eigenvalues
$\mu\in\sigma_d({^\lambda\!\Delta})$ survive for $\lambda=0$: the
eigenvalues $\mu<\nu_1$ become the discrete eigenvalues of $\Delta$,
while the eigenvalues $\mu>\nu_1$ become the embedded non-threshold
eigenvalues of $\Delta$. Only the Dirichlet Laplacian may have
discrete eigenvalues. Otherwise  $\nu_1=0$, and  all eigenvalues of
$\Delta$ are embedded into the absolutely continuous spectrum. In
view of the fact that  any non-threshold point $\mu$ can be
separated from $\sigma_{ess}({^\lambda \!\Delta})$ by a small
variation of $\arg(1+\lambda)$,  the set $\sigma_d({^\lambda
\!\Delta})$ (and therefore the set of all eigenvalues of $\Delta$)
has no accumulation points, except possibly for the thresholds
$\nu_1,\nu_2,\dots$. (In the companion paper~\cite{KalvinIII} we
show that the eigenvalues of $\Delta$ are of finite multiplicity and
can accumulate at the thresholds only from below.) Suitable examples
of manifolds, for which the eigenvalues of $\Delta$ do accumulate at
thresholds, can be found e.g. in~\cite{Edward, KalvinIII}.
  By definition, all discrete non-real
eigenvalues of ${^\lambda \!\Delta}$ are resonances of the
Laplacian. By the assertion~{\it 5} the resonances are characterized
by the pair $\{\Delta, \mathcal A\}$. They are identified with the
complex poles of the meromorphic continuation to a Riemann surface
of all resolvent matrix elements
$\mu\mapsto\bigl((\Delta-\mu)^{-1}F,G\bigr)$ with $F,G\in\mathcal
A$. The real poles correspond to the non-threshold eigenvalues of
the Laplacian. The embedded eigenvalues are known to be very
unstable, under rather weak perturbations they shift from the real
axis and become resonances, e.g.~\cite{Aslan}. Readers might have
noticed a certain analogy between the situation we described above
and the one known from the theory of resonances for N-body quantum
scattering problem e.g.~\cite{Hislop Sigal,Hunziker}.  As shown
in~\cite{MV1,MV2}, there is also a certain connection between N-body
quantum scattering and spectral theory of the Laplacian on symmetric
spaces.

Let us remark here that the assumption $\alpha<\pi/4$ in
Theorem~\ref{T1} is made for simplicity only. By taking $F$ and $G$
from different sets of analytic vectors, our results extend to all
$\alpha<\pi/2$. However, $\alpha=\pi/2$ is a substantial limit, as
in contrast to ${^{\lambda}\!\Delta}$ with $|\lambda|<1$ the
deformations ${^{\lambda}\!\Delta}$ of the Laplacian with
$\arg(1+\lambda)=\pm\pi/2$ are not elliptic operators.

\section{Geometry of the complex scaling}\label{s3}
In this section we deform the Riemannian global inner product on
$(\mathcal M,\mathsf g)$ by means of the complex scaling. In terms
of the deformed global inner product we introduce a sesquilinear
quadratic form $\mathsf q_\lambda$ associated with the unbounded
nonselfadjoint operator ${^\lambda\!\Delta}$ in the Hilbert space
$L^2(\mathcal M)$.

Let  $\mathrm T'^*\Bbb S_\alpha$ be the holomorphic cotangent bundle
$\{(z, c\,dz):z\in\Bbb S_\alpha, c\in\Bbb C\}$ of the sector $\Bbb
S_\alpha=\{z\in\Bbb C:|\arg z|<\alpha<\pi/4\}$, where $dz=d\Re
z+id\Im z$. Consider the tensor field
\begin{equation}\label{TF}
 \mathfrak g_0 dz\otimes dz +2\mathfrak g_1\otimes dz+\mathfrak g_2\in C^\infty({\mathrm T}'^*\Bbb
S_\alpha\otimes{\mathrm T}^*\Omega )^{\otimes 2}
\end{equation}
with the analytic coefficients $\Bbb S_\alpha\ni z\mapsto \mathfrak
g_k(z)\in C^\infty \mathrm T^*\Omega^{\otimes k}$, cf.
Definition~\ref{ACE}.

\begin{figure}
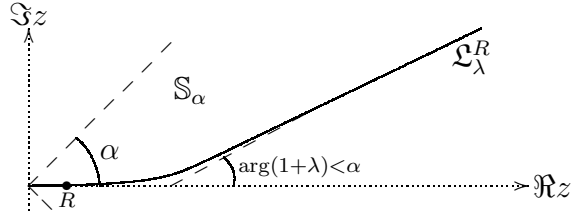
 \[ \xy0;/r.15pc/:
{\ar@{.>} (-30,0);(75,0)};(80,0)*{\Re z};
(-30,0);(0,30)**\dir{--};(-30,0);(-25,-5)**\dir{--};
(-22,0)*{\scriptstyle\bullet};(-22,-3)*{\scriptstyle R};
(-15,0);(-20,10)**\crv{(-16,7)};(-13,7)*{\alpha};(4,20)*{\Bbb
S_\alpha};
{\ar@{.>} (-30,-5);(-30,33)}; (-30,36)*{ \Im z};
(-30,0);(65,33)**\crv{(-5,0)&(5,4)&(15,9)&(36,19)};
(0,0);(28,15)**\dir{--};
(11,6);(13,0)**\crv{(14,3)};(27,4)*{\scriptstyle\arg(1+\lambda)<\alpha};
(63,27)*{ \mathfrak L^R_\lambda};
\endxy\]
\caption{The curve $\mathfrak L^R_\lambda=\{z\in\Bbb C: z=x+\lambda
\mathsf s_R(x),x\in\Bbb R_+ \}$ with $\lambda\in\mathcal
D_\alpha$.}\label{fig+}
\end{figure}

Recall that   $\mathsf s_R(x)=\mathsf s(x-R)$ is the scaling
function, where $\mathsf s\in C^\infty(\Bbb R)$ has the
properties~\eqref{ab}, and $R>0$ is a sufficiently large number. For
all values of the scaling parameter $\lambda$ in the
disk~\eqref{disk} the complex scaling $\Bbb R_+\ni x\mapsto
x+\lambda \mathsf s_R(x)$, see Fig.~\ref{fig+}, defines the
embedding
\begin{equation}\label{emb}
\mathrm T^*\Bbb R_+\ni\{x, a\,dx\}\mapsto \{x+\lambda \mathsf
s_R(x), a(1+\lambda \mathsf s'_R(x))^{-1} d z\}\in \mathrm T'^*\Bbb
S_\alpha,
\end{equation}
where $|1+\lambda \mathsf s'_R(x)|>1-1/\sqrt{2}$.
As before, we identify the bundle $\mathrm T^*\Pi$ with the product
$\mathrm T^*\Bbb R_+\otimes \mathrm T^*\Omega$ via the  product
structure on $\Pi=\Bbb R_+\times\Omega$. The embedding~\eqref{emb}
together with~\eqref{TF} induces the tensor field
\begin{equation}\label{TFL}
\begin{aligned}
\mathsf g_\lambda\!\upharpoonright_\Pi=\mathfrak g^R_{0,\lambda}  dx\otimes dx +2\mathfrak g^R_{1,\lambda} \otimes dx +\mathfrak g^R_{2,\lambda}\in C^\infty {\mathrm T}^* \Pi^{\otimes 2},\\
\mathfrak g^R_{k,\lambda}(x)=(1+\lambda \mathsf
s'_R(x))^{2-k}\mathfrak g_{k}(x+\lambda \mathsf s_R(x)),
\end{aligned}
\end{equation}
where $\mathfrak g^R_{k,\lambda}(x)\in C^\infty \mathrm
T^*\Omega^{\otimes k}$ are  smooth in $x\in\Bbb R_+$ and analytic in
$\lambda\in\mathcal D_\alpha$ coefficients.
 Since $\supp
\mathsf s_R\cap (0,R)=\varnothing$, the equality $\mathsf
g_\lambda\!\upharpoonright_{(0,R)\times\Omega}=\mathsf
g\!\upharpoonright_{(0,R)\times\Omega}$ holds for all
$\lambda\in\mathcal D_\alpha$, cf.~\eqref{split} and~\eqref{TFL}.
Thanks to this we can smoothly extend $\mathsf
g_\lambda\!\upharpoonright_{\Pi}$  to $\mathcal M$ by setting
$\mathsf g_\lambda\!\upharpoonright_{\mathcal M\setminus\Pi}=\mathsf
g\!\upharpoonright_{\mathcal M\setminus\Pi}$. As a result we obtain
an analytic function
$$
\mathcal D_\alpha\ni\lambda\mapsto \mathsf g_\lambda \in C^\infty
\mathrm T^*\mathcal M^{\otimes 2}.
$$

We consider the tensor field $\mathsf g_\lambda$  with
$\lambda\in\mathcal D_\alpha$ as a deformation of the metric
$\mathsf g$ on $\mathcal M$ by means of the complex scaling.
Clearly, $\mathsf g_0\equiv\mathsf g$. Moreover, if
$\lambda\in\mathcal D_\alpha$ is real, then $\mathsf g_\lambda$
coincides with the metric  introduced  in Section~\ref{s2} as the
pullback $\varkappa_\lambda^*\mathsf g$.   By analyticity in
$\lambda$ we conclude that $\mathsf g_\lambda$ is a symmetric tensor
field. The Schwarz reflection principle gives $\overline{\mathsf
g_\lambda}=\mathsf g_{\overline{\lambda}}$, where the bar stands for
the complex conjugation. It must be stressed that the  tensor field
$\mathsf g_\lambda$ with $\lambda\neq 0$ depends on a sufficiently
large number $R$, however we do not indicate this for brevity of
notations. On the next step we clarify the behaviour of the tensor
field $\mathsf g_\lambda$ at infinity, relying on
Definition~\ref{ACE} of an axial analytic asymptotically cylindrical
end.

By Definition~\ref{ACE} the analytic functions  $\Bbb S_\alpha\ni
z\mapsto \mathfrak g_k(z)$ uniformly tend to some limits in the norm
$\|\cdot\|_{\mathfrak e}$ of $C^1\mathrm T^*\Omega^{\otimes k}$ as
$|z|\to\infty$. Hence the values $\|\partial^q_z\mathfrak
g_k(z)\|_{\mathfrak e}$ with $q=1,2,\dots$ uniformly tend to zero as
$z$ tends to infinity in the smaller sector $\{z\in\Bbb C:
|\arg(z-1)|<\alpha\}\subset \Bbb S_\alpha$; here
$\partial_z=(\partial_{\Re z}-i\partial_{\Im z})/2$ is the complex
derivative. Taking into account the properties of the scaling
function $\mathsf s_R$, we conclude that
\begin{equation}\label{1}
\begin{aligned}
\|\partial_x^q(\mathfrak g^R_{0,\lambda}(x)-(1+\lambda)^2) \|_{\mathfrak e}+\|\partial_x^q \mathfrak g^R_{1,\lambda}(x)\|_{\mathfrak e}+\|\partial_x^q(\mathfrak g^R_{2,\lambda}(x)-\mathfrak h)\|_{\mathfrak e}\leq c_q(x),\\
c_q(x)\to 0\text{ as }  x\to +\infty,\quad q=0,1,\dots,\quad
\lambda\in\mathcal D_\alpha,
\end{aligned}
\end{equation}
where the coefficients $\mathfrak g^R_{k,\lambda}$ are the same as
in~\eqref{TFL}, and $\partial_x$ is the real derivative. Thus, in
the sense of~\eqref{TFL} and~\eqref{1},  the tensor field $\mathsf
g_\lambda$  stabilizes at infinity to the  tensor field
$(1+\lambda)^2\,dx\otimes dx+\mathfrak h$ on $\Pi$ uniformly in
$\lambda\in\mathcal D_\alpha$. Observe also that $|x+\lambda\mathsf
s_R(x)|\geq R$ for $x\geq R$, and hence
\begin{equation}\label{2}
\begin{aligned}
\|\partial_x^q(\mathfrak g^R_{0,\lambda}(x)-(1+ \lambda\mathsf
s'_R(x))^2) & \|_{\mathfrak e}+\|\partial_x^q \mathfrak
g^R_{1,\lambda}(x)\|_{\mathfrak e}
\\
&+\|\partial_x^q(\mathfrak g^R_{2,\lambda}(x)-\mathfrak h)\|_{\mathfrak e}\leq C_q(R),\ x\geq R;\\
C_q(R)\to 0 &\text{ as } R\to+\infty, \quad q=0,1,\dots,\quad
\lambda\in\mathcal D_\alpha.
\end{aligned}
\end{equation}
Since $R$ is supposed to be sufficiently large,~\eqref{TFL}
and~\eqref{2} imply that on the subset $[R,\infty)\times\Omega$ of
$\Pi$ the tensor field  $\mathsf g_\lambda$ is close to
$$
\overset{\infty}{\mathsf g}_\lambda:=(1+\lambda\mathsf
s_R')^2\,dx\otimes dx+\mathfrak h\in C^\infty {\mathrm T}^*
\Pi^{\otimes 2}.
$$
 In particular,   $\mathsf g_\lambda\!\upharpoonright_\Pi$ is exactly equal to $\overset{\infty}{\mathsf g}_\lambda$, if $(\Pi,\mathsf g\!\upharpoonright_{\Pi})$ is a cylindrical end; i.e. if $\mathsf g\!\upharpoonright_{\Pi}\equiv\overset{\infty}{\mathsf g}$, where $\overset{\infty}{\mathsf g}=dx\otimes dx+\mathfrak h$ is the product metric on $\Pi$.

In what follows we will often work in local coordinates on $\Omega$.
Recall from Section~\ref{s0} that $\{\mathscr U_j,\kappa_j\}$ is a
finite atlas on $\Omega$, and $y\in\Bbb R^n$ is a system of local
coordinates in $\mathscr U_j$. Let $\mathbf g(x,y)$ be the matrix
corresponding to the representation of the metric $\mathsf g$ in the
coordinates $(x,y)\in \Bbb R_+\times \kappa_j[\Omega\cap\mathscr
U_j]$ on $\Pi$.  Let us use the Einstein summation convention for
the indexes varying from $1$ to $n$. Then by virtue of~\eqref{TFL}
the symmetric tensor field $\mathsf g_\lambda$  has the local
coordinate representation
$$
\mathsf g_\lambda=\mathbf g_{\lambda,00}\,dx\otimes dx+2\mathbf
g_{\lambda,0\ell}\,dy_\ell\otimes dx+\mathbf g_{\lambda,\ell
m}\,dy_\ell\otimes dy_m,
$$
where the matrix  $\mathbf g_\lambda(x,y)=\{\mathbf g_{\lambda,\ell
m}(x,y)\}_{\ell,m =0}^n$ is given by the equality
\begin{equation}\label{metric}
\mathbf g_\lambda(x,y)=\diag\left\{1+\lambda \mathsf
s'_R(x),\operatorname{Id} \right\}\mathbf g(x+\lambda \mathsf
s_R(x),y)\diag\left\{1+\lambda \mathsf s'_R(x),\operatorname{Id}
\right\}.
\end{equation}
Here  $\operatorname{Id}$ is the $n\times n$-identity matrix, and
$\mathbf g(x+\lambda \mathsf s_R(x),y)$  stands for the value of the
analytic function $\Bbb S_\alpha\ni z\mapsto \mathbf g(z,y)$ at the
point $z=x+\lambda \mathsf s_R(x)$; cf.~Definition~\ref{R1}.
Clearly, $\mathcal D_\alpha\ni\lambda\mapsto \mathbf g_\lambda(x,y)$
is an analytic function, whose  values are complex symmetric
matrices. Moreover,  $\overline{\mathbf g_\lambda(x,y)}=\mathbf
g_{\overline{\lambda}}(x,y)$.

To the  representation of the tensor field $\overset{\infty}{\mathsf
g}_\lambda$ with $\lambda\in\mathcal D_\alpha$  in the coordinates
$(x,y)$ there corresponds the invertible matrix
\begin{equation}\label{h lambda}
\overset{\infty}{\mathbf g}_\lambda(x,y)=\diag\{(1+\lambda\mathsf
s'_R(x))^2,\mathbf h(y)\},
\end{equation}
where $\mathbf h(y)$ is the matrix coordinate representation of the
metric $\mathfrak h$ on $\Omega$. Note that the matrices $\mathbf
g_\lambda$ and $\overset{\infty}{\mathbf g}_\lambda$ both depend on
a sufficiently large number $R$, however we do not indicate this for
brevity of notations.

\begin{lem}\label{sss}
\begin{itemize}

\item[1.] Definition~\ref{ACE} and Definition~\ref{R1} of an axial analytic asymptotically cylindrical end are equivalent.

\item[2.] Let $R>0$ be a  sufficiently large number. Then the matrix $\mathbf g_\lambda(x,y)$ is invertible for all $\lambda\in\mathcal D_\alpha$ and $(x,y)\in \Bbb R_+\times \kappa_j[\Omega\cap\mathscr U_j]$, and the inverse matrix $\mathbf g^{-1}_\lambda(x,y)$ meets the estimate
\begin{equation}\label{3}
\sum_{q+|r|\leq 1}\|\partial_x^q\partial_y^r(\mathbf
g^{-1}_\lambda(x,y)-\overset{\infty}{\mathbf g}\vphantom{\mathbf
g}^{-1}_\lambda(x,y))\|_{2}\leq C(R) \text{ for } x\geq R,
\end{equation}
 where $C(R)$ tends to zero uniformly in $\lambda\in\mathcal D_\alpha$, $x\geq R$,  and $y\in  \kappa_j[\Omega\cap\mathscr U_j]$ as $R\to+\infty$. Moreover,
\begin{equation}\label{4}
\sum_{q+|r|\leq 1}\|\partial_x^q\partial_y^r(\mathbf
g^{-1}_\lambda(x,y)-\diag\{(1+\lambda)^{-2},\mathbf
h^{-1}(y)\})\|_{2}\to 0 \text{ as } x\to+\infty,
\end{equation}
uniformly in $\lambda\in\mathcal D_\alpha$ and $y\in
\kappa_j[\Omega\cap\mathscr U_j]$.
\end{itemize}
\end{lem}
\begin{pf} {\it 1.} The proof is straightforward. For brevity we assume that the atlas $\{\mathscr U_j,\kappa_j\}$ on $\Omega$ consists of only one coordinate neighborhood $\{\mathscr U,\kappa\}$. Then
$\mathfrak g_0(z,\mathrm y)=\mathbf g_{00}(z,y)$, $\mathfrak
g_1(z,\mathrm y)=\mathbf g_{0\ell}(z, y)dy_\ell$, and $\mathfrak
g_2(z,\mathrm y)=\mathbf g_{\ell m}(z,y)d y_\ell\otimes dy_m$, where
$y=\kappa(\mathrm y)$ and $z\in\Bbb S_\alpha$. Hence the first
condition in Definition~\ref{ACE} is valid, if and only if  $\Bbb
S_\alpha\ni z\mapsto \mathbf g_{\ell m}(z,\cdot)\in
C^\infty(\kappa[\Omega\cap\mathscr U])$, where $\ell,m=0,1,\dots,n$,
are analytic functions. The metric matrix $\mathbf g(x,y)$ with
$x\in \Bbb R_+$ is symmetric. Therefore $\mathbf g_{\ell
m}(z,y)=\mathbf g_{ m \ell}(z,y)$ by analyticity in $z\in \Bbb
S_\alpha$.

Let $D$ be  the Levi-Civita connection on  $\Omega$ associated with
a metric $\mathfrak e$.  We have
$$
\begin{aligned}
D\mathfrak g_0=(\partial_{y_\ell}\mathbf g_{00})\,d y_{\ell},\quad D\mathfrak g_1=(\partial_{y_m}\mathbf g_{0\ell} ) d y_m\otimes d y_\ell + \mathbf g_{0\ell} Ddy_\ell,\\
D\mathfrak g_2= (\partial_{y_j}\mathbf g_{\ell m} ) dy_j\otimes
dy_\ell\otimes d y_m+\mathbf g_{\ell m} D( dy_\ell\otimes d y_m),
\end{aligned}
$$
where $D dy_\ell=-\Gamma^\ell_{jm} dy_j\otimes d y_m$ with the
Christoffel symbols $\Gamma^\ell_{jm}$ related to $\mathfrak e$, and
$D( dy_\ell\otimes d y_m)=(D dy_\ell)\otimes d y_m+ dy_\ell\otimes
Dd y_m$.

As before, we denote by $|\cdot|_{\mathfrak e}(\mathrm y)$ the norm
induced by $\mathfrak e$ in  $\Bbb C\mathrm T_{\mathrm
y}^*\Omega^{\otimes k}$. Since the manifold $(\Omega,\mathfrak e)$
is  compact, the relation $|\xi_\ell d y_\ell|_{\mathfrak e}(\mathrm
y)\asymp |\xi|$  holds, i.e. for some $\epsilon>0$ and all $\mathrm
y\in\Omega$ and  $\xi\in \Bbb C^n$ we have $\epsilon |\xi|\leq
|\xi_\ell d y_\ell|_{\mathfrak e}(\mathrm y)\leq |\xi|/\epsilon$. As
a consequence, after simple manipulations, based on the
Cauchy-Schwarz inequality, we get
$$
\|\mathfrak g_{0}(z)-1\|_{\mathfrak e}\asymp\max_{y}\Bigl( |\mathbf
g_{00}(z,y)-1|^2 +\sum_{\ell=1}^n|\partial_{y_\ell}\mathbf
g_{00}(z,y)|^2\Bigr)^{1/2},
$$
$$
\|\mathfrak g_1(z)\|_{\mathfrak e}\asymp\max_y
\Bigl(\sum_{\ell=1}^n|\mathbf g_{0\ell}(z,y)|^2+\sum_{\ell,m
=1}^n|\partial_{y_m}\mathbf g_{0\ell}(z,y)+\Gamma^j_{\ell
m}(y)\mathbf g_{0 j}(z,y) |^2 \Bigr)^{1/2},
$$
$$
\begin{aligned}
\|\mathfrak g_2(z)-& \mathfrak h\|_{\mathfrak e}\asymp\max_y\Bigl(\sum_{\ell,m=1}^n |\mathbf g_{\ell m}(z,y)-\mathbf h_{\ell m}(y)|^2\\
 & +\sum_{\ell,m,j=1}^n|\partial_{y_j}(\mathbf g_{\ell m}(y,z)-\mathbf h_{\ell m}(y))+(\mathbf g_{k  m}(z,y)-\mathbf h_{k m}(y))\Gamma^k_{\ell j}(y)|^2\Bigr)^{1/2}.
\end{aligned}
$$
These relations together with the identity $\overset{\infty}{\mathbf
g}(y)=\diag\{1,\mathbf h(y)\}$ show that the second condition in
Definition~\ref{ACE} is equivalent to the stabilization
condition~\eqref{stab}.

{\it 2.} For all $z\in\Bbb S_\alpha$ with sufficiently large $|z|$
the matrix $\mathbf g(z,y)$ is invertible, because the metric matrix
$\overset{\infty}{\mathbf g}(y)$  is invertible, and the norm
$\|\mathbf g(z,y)-\overset{\infty}{\mathbf g}(y)\|_{2}$ is small by
the stabilization condition~\eqref{stab}. As $R>0$ is a sufficiently
large number, and $|x+\lambda \mathsf s_R(x)|\geq R$ for $x\geq R$,
the equality~\eqref{metric} and the stabilization
condition~\eqref{stab} imply that the matrix  $\mathbf
g_\lambda(x,y)$ is invertible for all $x\geq R$,
$y\in\kappa_j[\Omega\cap\mathscr U_j]$, and $\lambda\in\mathcal
D_\alpha$. On the other hand, for all $x<R$ the matrix  $\mathbf
g_\lambda(x,y)$ coincides with the matrix $\mathbf g(x,y)$ of the
metric $\mathsf g$, because $\mathsf s_R(x)=0$. Therefore the matrix
$\mathbf g_\lambda(x,y)$ is invertible for all $x\in\Bbb R_+$,
$y\in\kappa_j[\Omega\cap\mathscr U_j]$, and $\lambda\in\mathcal
D_\alpha$.

Similarly to the proof of the first assertion, from the
relations~\eqref{2}  we obtain
\begin{equation}\label{5}
\begin{aligned}
\sum_{q+|r|\leq 1}\|\partial_x^q\partial_y^r(\mathbf g_\lambda(x,y)-\overset{\infty}{\mathbf g}_\lambda(x,y))\|_{2}\leq c(R) \text{ for } x\geq R;\\
c(R)\to 0\text{ as } R\to+\infty,
\end{aligned}
\end{equation}
where the constant $c(R)$ is independent of $x\geq R$,
$y\in\kappa_j[\Omega\cap\mathscr U_j]$, and $\lambda\in\mathcal
D_\alpha$. Note that~\eqref{5} can also be derived as a consequence
of the stabilization condition~\eqref{stab} and the
equalities~\eqref{metric},~\eqref{h lambda}. The identity
$$
\partial_x^q\partial_y^r(\mathbf g^{-1}_\lambda -\overset{\infty}{\mathbf g}\vphantom{\mathbf g}^{-1}_\lambda)=
-\mathbf g^{-1}_\lambda \bigl(\partial_x^q\partial_y^r\mathbf
g_\lambda\bigr)\mathbf g^{-1}_\lambda+\overset{\infty}{\mathbf
g}\vphantom{\mathbf g}^{-1}_\lambda\bigl(\partial_x^q\partial_y^r
\overset{\infty}{\mathbf g}_\lambda\bigr)\overset{\infty}{\mathbf
g}\vphantom{\mathbf g}^{-1}_\lambda,\quad q+|r|=1,
$$
together with~\eqref{5} gives the estimate~\eqref{3}, where $C(R)$
tends to zero uniformly in $\lambda\in\mathcal D_\alpha$ and
$(x,y)\in [R,\infty)\times \kappa_j[\Omega\cap\mathscr U_j]$ as
$R\to+\infty$.

Similarly, the property~\eqref{4} is a consequence of~\eqref{1} (or,
equivalently, it is a consequence of the stabilization
condition~\eqref{stab}, the equality~\eqref{metric}, and  properties
of the scaling function $\mathsf s_R$). \qed\end{pf}

Recall that with every real $\lambda\in\mathcal D_\alpha$ we
associate the Laplacian ${^\lambda\!\Delta}$ on the Riemannian
manifold $(\mathcal M,\mathsf g_\lambda)$, and also the operator
${^\lambda\partial_\nu}$ of the Neumann boundary condition, if
$\partial\mathcal M\neq\varnothing$. Observe that
${^\lambda\!\Delta}\equiv {^0\!\Delta}$ and
${^\lambda\!\partial_\nu\equiv {^0\!\partial_\nu}}$ on $\mathcal
M\setminus(\supp \mathsf s_R\times\Omega)$, because $\mathsf
g_\lambda\equiv\mathsf g$ on this set. At the same time,  in the
local coordinates $(x,y)$ on $\Pi\supset \supp \mathsf
s_R\times\Omega$ we have
\begin{equation}\label{LR}
 \begin{aligned}
 {^\lambda\!\Delta}&=-\frac {1} {\sqrt{\det \mathbf
 g_\lambda }}\nabla_{xy}\cdot \sqrt{\det \mathbf
 g_\lambda }\,\,\mathbf g^{-1}_\lambda
 \nabla_{xy},
 \\
{^\lambda\!\partial_\nu}&=\bigl(0,\dots,0,1/\sqrt{\mathbf
g^{-1}_{\lambda,nn}}\bigr)\mathbf
g^{-1}_\lambda \upharpoonright_{y_n=0}\nabla_{xy}\quad\text{ if }\quad  \mathscr U_j\cap\partial\Omega\neq \varnothing,\\
\end{aligned}
\end{equation}
where  $\lambda\in\mathcal D_\alpha\cap\Bbb R$, the matrix $\mathbf
g_\lambda$ is given in~\eqref{metric}, and
$\nabla_{xy}\equiv(\partial_x,\partial_{y_1},\dots,\partial_{y_n})^\intercal$.
Due to the properties of  $\mathbf g_\lambda$, the coefficients
$a^{rq}_\lambda(x,y)$ of the differential operators~\eqref{LR},
written in the form $\sum a_\lambda^{rq}\partial_x^r\partial_y^q$,
are analytic functions of   $\lambda\in\mathcal D_\alpha$. Hence
the operators ${^\lambda\!\Delta}: C_c^\infty(\mathcal M)\to
C_c^\infty(\mathcal M)$ and
${^\lambda\partial_\nu}:C_c^\infty(\mathcal M)\to
C_c^\infty(\partial\mathcal M) $ extend by analyticity from real to
all $\lambda$ in the disk $\mathcal D_\alpha$.

In general, the operator ${^\lambda\partial_\nu}$ and the core
$\mathbf C({^\lambda\!\Delta})$ of the deformed Neumann Laplacian
${^\lambda\!\Delta}$ depend on $\lambda\in\mathcal D_\alpha$; this
is easily seen from~\eqref{LR}, Definition~\ref{d1}, and
illustrative examples in Section~\ref{s0}. For this reason (and also
because it is convenient to introduce m-sectorial operators via
their quadratic forms) our study of the analytic family of unbounded
operators $\mathcal D_\alpha\ni\lambda\mapsto{^\lambda\!\Delta}$
will be based on analysis of the corresponding family of quadratic
forms. The quadratic form of the operator ${^\lambda\!\Delta}$ will
be introduced in terms of the sesquilinear form $\mathsf
g_\lambda^p[\cdot,\cdot]$ on the complexified cotangent space $\Bbb
C\mathrm T_p^*\mathcal M$, induced  by the tensor field $\mathsf
g_\lambda$ for every  $p\in\mathcal M$. Clearly, the tensor field
$\mathsf g_\lambda\in C^\infty\mathrm T^*\mathcal M^{\otimes 2}$
naturally defines the sesquilinear form $\mathsf
g_\lambda^p[\cdot,\cdot]$ on  the complexified tangent space $\Bbb
C\mathrm T_p\mathcal M$.
 On the next step we extend $\mathsf g_\lambda^p[\cdot,\cdot]$ to $\Bbb C\mathrm T_p^*\mathcal M$.

Observe that for all $\lambda\in\mathcal D_\alpha$ the tensor field
$\mathsf g_\lambda$ is non-degenerate, because  $\mathsf g_\lambda$
coincides with the metric $\mathsf g$ on $\mathcal M\setminus\Pi$,
and in every coordinate neighborhood $\mathscr U_j$ on $\Omega$ the
matrix $\mathbf g_\lambda(x,y)$ is invertible for all
$\lambda\in\mathcal D_\alpha$ and $x\in \Bbb R_+$ by
Lemma~\ref{sss}. As is well known, a Riemannian metric induces a
musical isomorphism between the tangent and cotangent bundles,
e.g.~\cite{Gallot}. In a similar way the non-degenerate tensor field
$\mathsf g_\lambda\!\upharpoonright_{\Pi}$ induces a fiber
isomorphism between the complexified  bundles $\Bbb  C{\mathrm
T}\Pi$ and $\Bbb  C{\mathrm T}^*\Pi$. Indeed, let $\xi\in {\Bbb
C{\mathrm T}^*_{p}\Pi}(={\mathrm T}^*_{p}\Pi\otimes \Bbb C)$ and
$\zeta,\eta\in {\Bbb  C{\mathrm T}_{p}\Pi}(={\mathrm
T}_{p}\Pi\otimes \Bbb C)$. In the local coordinates  we have
\begin{equation}\label{lcpr}
\xi=\xi_0\, d x+\xi_1\,d{y_1}+\cdots+\xi_{n}\,d{y_n}, \quad
\zeta=\zeta_0\,
\partial_x+\zeta_1\partial_{y_1}+\cdots+\zeta_{n}\partial_{y_n},
\end{equation}
and a similar expression for $\eta$, where $\xi_j$, $\zeta_j$, and
$\eta_j$ are complex coefficients. Since the tensor field $\mathsf
g_\lambda\!\upharpoonright_{\Pi}$ is non-degenerate, for any $\xi$
there exists a unique $\zeta$, such that for all $\eta$ the
following equalities hold
\begin{equation}\label{flat}
 \xi\overline{\eta}=\xi_m\bar{\eta}_m={\mathbf g_{\lambda,\ell m}(x,y)} \zeta_\ell \bar{\eta}_m=\mathsf g_\lambda^p [\zeta,\eta].
\end{equation}
Therefore  $\mathsf g_\lambda\!\upharpoonright_{\Pi}$ induces a
musical  isomorphism ${^\lambda\flat}:{\Bbb  C{\mathrm
T}\Pi}\to{\Bbb C{\mathrm T}^*\Pi}$, such that in each fiber we have
${^\lambda\flat}\zeta=\xi$ with $\xi_m=\mathbf g_{\lambda,\ell
m}(x,y) \zeta_\ell $. The operator inverse  to $^\lambda\flat$ will
be denoted by $^\lambda\sharp$. As a result, the non-degenerate
tensor field $\mathsf g_\lambda$ induces a musical isomorphism
$^\lambda\sharp:\Bbb C\mathrm T^*\mathcal M\to \Bbb C \mathrm
T\mathcal  M$ between the complexified tangent and cotangent
bundles. On $\Pi$ this isomorphism coincides with the constructed
above isomorphism $^\lambda\sharp:\Bbb C\mathrm T^*\Pi\to \Bbb C
\mathrm T\Pi$, and on $\mathcal M\setminus[R,\infty)\times\Omega$ it
coincides with the complexified Riemannian musical isomorphism
$\sharp:\Bbb C\mathrm T^*\mathcal M\to\Bbb C\mathrm T\mathcal M$
induced  by the metric $\mathsf g$.

We extend the sesquilinear form ${\mathsf g_\lambda^p}
[\cdot,\cdot]$ to the pairs $(\xi,\omega)\in {\Bbb  C{\mathrm
T}^*_{p}\mathcal M}\times {\Bbb C{\mathrm T}^*_{p}\mathcal M}$ by
the equality
\begin{equation}\label{mus iso}
{\mathsf g_\lambda^p} [\xi,\omega]={\mathsf g_\lambda^p}
[{^\lambda\sharp}\,\xi,{^{\overline{\lambda}}\sharp}\,\omega], \quad
\lambda\in\mathcal D_\alpha.
\end{equation}
If $\lambda\in\mathcal D_\alpha$ is real, then ${\mathsf
g_\lambda^p} [\cdot,\cdot]$ is the positive Hermitian form
corresponding to the Riemannian metric $\mathsf g_\lambda$ on
$\mathcal M$. In particular, ${\mathsf g_0^p} [\cdot,\cdot]\equiv
\mathsf g^p [\cdot,\cdot]$. For a non-real $\lambda\in\mathcal
D_\alpha$ we have $\overline{{\mathsf g_\lambda^p}
[\xi,\omega]}={\mathsf g_{\overline \lambda}^p} [\omega,\xi]$, and
the form ${\mathsf g_\lambda^p} [\cdot,\cdot]$ is not Hermitian.

 Similarly, ${\mathsf g_\lambda}$ with $\lambda\in\mathcal D_\alpha$ induces a  sesquilinear form on each tensor power ${\Bbb  C {\mathrm
T}_{p}\mathcal M}^{\otimes k}$ and  $\Bbb C\mathrm T^*_p\mathcal
M^{\otimes k}$. Since we study the Laplacian on functions, for our
aims it suffices to consider  ${\mathsf g_\lambda^p} [\cdot,\cdot]$
on the differential one-forms.
\begin{lem}\label{l1} Assume that the parameter $R$ is sufficiently large. Then there exist $\vartheta< \pi/2$ and $\delta >0$, such that for all $p\in \mathcal M$ and $\lambda\in\mathcal D_\alpha$  we have
 $$
 |\arg {\mathsf g_\lambda^p} [\xi,\xi]|\leq \vartheta, \quad \delta {\mathsf g^p} [\xi,\xi]\leq\Re{\mathsf g_\lambda^p} [\xi,\xi]\leq \delta^{-1} {\mathsf g^p} [\xi,\xi] \quad \forall \xi\in \Bbb C\mathrm T^*_p \mathcal M.$$
  In other words,  on the differential one-forms the sesquilinear  quadratic form ${\mathsf g_\lambda^p} [ \cdot,\cdot ]$ is sectorial and relatively bounded, the  sector and the  bounds  are independent of $p$ and  $\lambda$.
\end{lem}
\begin{pf} By construction of the tensor field~$\mathsf g_\lambda$,  for all $p\in\mathcal M\setminus\{(R,\infty)\times\Omega\}$ and $\lambda\in\mathcal D_\alpha$ we have $\mathsf g_\lambda^p=\mathsf g^p$. Here $\mathsf g$ is the Riemannian metric on $\mathcal M$, and therefore $\mathsf g^p[\xi,\xi]\geq 0$.
 Let $p\in(R,\infty)\times\Omega$ and  $\zeta={^\lambda\sharp}\,\xi$. In the local coordinates we have the representations~\eqref{lcpr}. From~\eqref{mus iso}
and~\eqref{flat} we conclude that $\mathbf g_{\lambda,\ell
m}(x,y)\zeta_m=\xi_\ell$. Therefore, by virtue of the fact that
$\overline{\mathbf g_\lambda}=\mathbf g_{\overline{\lambda}}$, the
expression $\mathsf g^p_\lambda[ {^\lambda\sharp}\,\xi,
{^{\overline{\lambda}}\sharp}\,\xi]$ on the one-forms $\xi\in\Bbb
C{\mathrm T}^*_{p}\Pi$ can be written as $\overline{\xi}\cdot\mathbf
g_\lambda^{-1}(x,y)  \xi$, where we identify the one-form $\xi$ with
the vector of coefficients
$\xi=(\xi_0,\xi_1,\dots,\xi_n)^\intercal\in \Bbb C^{n+1}$,
cf.~\eqref{lcpr}.
 It remains to
show that for all $\lambda\in\mathcal D_\alpha$ and  $ x>R$, where
$R>0$ is   sufficiently large,  the set
\begin{equation}\label{num}
\begin{aligned}
\bigl\{z\in\Bbb C: z=\overline{\xi}\cdot\mathbf g_\lambda^{-1}(x,y)
\xi,\ \xi\in \Bbb C^{n+1}\bigr\}
\end{aligned}
\end{equation}
is inside the sector  $\{z\in\Bbb C: |\arg z|\leq\vartheta\}$ of
angle $2\vartheta <\pi$, and the estimate
\begin{equation}\label{es}
\delta^{1/2} |\xi|^2\leq\Re (\overline{\xi}\cdot\mathbf
g_\lambda^{-1}(x,y)  \xi)\leq \delta^{-1/2}|\xi|^2,\quad \xi\in\Bbb
C^{n+1},
\end{equation}
 is valid for some $\delta>0$.

By Lemma~\ref{sss}  the uniform in $\lambda\in\mathcal D_\alpha$,
$x>R$, and $y\in\kappa_j[\Omega\cap\mathscr U_j]$ estimate
\begin{equation}\label{stab1}
\bigl|\overline{\xi}\cdot \bigl(\mathbf g^{-1}_\lambda(x,y) -
\overset{\infty}{\mathbf g}\vphantom{\mathbf
g}^{-1}_\lambda(x,y)\bigr) \xi\bigr|\leq \mathsf c(R)|\xi|^2
\end{equation}
is valid, where the constant $\mathsf  c(R)$ is independent of
$\xi\in\Bbb C^{n+1}$, and
 $\mathsf c(R)\to 0$ as $R\to\infty$. Hence the
estimate~\eqref{stab1} holds with a given arbitrarily  small
constant $\mathsf  c(R)>0$ as $R$ is sufficiently large.

Recall that  the matrix $\mathbf h(y)$ corresponds to the local
coordinate representation of the metric $\mathfrak h$ on $\Omega$.
By~\eqref{h lambda} we have
\begin{equation}\label{101}
\overline{\xi}\cdot \overset{\infty}{\mathbf g}\vphantom{\mathbf
g}^{-1}_\lambda(x,y) \xi =(1+\lambda \mathsf s'_R
(x))^{-2}{|\xi_0|^2}+\overline {\xi'}\cdot \mathbf h^{-1}(y) \xi',
\end{equation}
where $\xi'=(\xi_1,\dots,\xi_n)\in\Bbb C^{n}$. Since $\mathbf
h^{-1}(y)$ is a  symmetric positive definite matrix, we conclude
that
$$
\overline {\xi'}\cdot \mathbf h^{-1}(y) \xi'\geq C|\xi'|^2\quad
\forall \xi'\in\Bbb C^n
$$
with some $C>0$. Taking into account~\eqref{101}, and the
inequalities $|\lambda|<\sin\alpha$ and $0\leq \mathsf s'_R (x)\leq
1$, we arrive at the uniform in  $\lambda\in\mathcal D_\alpha$,
$R>0$, $x>R$, and $y\in\kappa_j[\Omega\cap\mathscr U_j]$
 estimates
\begin{equation}\label{eee-1}
\begin{aligned}
\bigl|\arg\bigl( \overline{\xi}  \cdot\overset{\infty}{\mathbf
g}\vphantom{\mathbf g}^{-1}_\lambda(x,y)
\xi\bigr)\bigr|<2\alpha<\pi/2,
\\
c |\xi|^2\leq\bigl| \overline{\xi}  \cdot\overset{\infty}{\mathbf
g}\vphantom{\mathbf g}^{-1}_\lambda(x,y) \xi\bigr|\leq c^{-1}
|\xi|^2,
\end{aligned}
\end{equation}
where $c>0$. These estimates together with the
estimate~\eqref{stab1}, where $\mathsf c (R)$ is sufficiently small,
imply that the set~\eqref{num} is inside the sector $|\arg z|\leq
\vartheta$  with $\vartheta=2\alpha+2\arcsin ({\mathsf
c(R)}/{2c})<\pi/2$, and the estimates~\eqref{es} are valid with
$\delta^{1/2}=\min\{c-\mathsf c (R),(1/c+\mathsf c (R))^{-1}\}$.
\qed\end{pf}

 Let  $\operatorname{dvol}_{\lambda}$ with real $\lambda\in\mathcal D_\alpha$ be the volume form on the  Riemannian manifold  $(\mathcal M,\mathsf g_\lambda)$. Introduce a  density $\varrho_\lambda$ on $\mathcal M$, such that
$\varrho_\lambda\operatorname{dvol}_{\lambda}=
\operatorname{dvol}_{0}$. It is clear that $\varrho_\lambda\equiv 1$
on $\mathcal M\setminus(\supp \mathsf s_R\times\Omega)$. At the same
time in the local coordinates on $\Pi$ we have
\begin{equation}\label{mu xy}
\varrho_\lambda=\sqrt{\det{\mathbf g_0}/\det{\mathbf
g_\lambda}},\quad \operatorname{dvol}_{\lambda}=\sqrt{\det \mathbf
g_\lambda}\,dx\wedge dy_1\wedge\dots\wedge dy_n,
\end{equation}
 where $\mathbf g_\lambda$ is the matrix~\eqref{metric}, and $\wedge$ is the wedge product. Hence $\varrho_\lambda\in C^\infty(\mathcal M)$ is an analytic function of $\lambda\in\mathcal D_\alpha$.
Note that the second assertion of Lemma~\ref{sss} gives the bounds
$0<c_1\leq|\varrho_\lambda(p)|\leq c_2$, where $c_1$ and $c_2$ are
independent of $p\in\mathcal M$ and $\lambda\in\mathcal D_\alpha$.
We introduce the deformed volume form
\begin{equation}\label{vol}
\operatorname{dvol}_{\lambda}:=\frac 1 {\varrho_\lambda}
\operatorname{dvol}_{0},\quad \lambda\in\mathcal D_\alpha,
\end{equation}
and the  deformed global inner product
\begin{equation}\label{form}
(\xi,\omega)_\lambda=\int_{\mathcal M} \mathsf g_\lambda[\xi,\omega]
\,\operatorname{dvol}_{\lambda}, \quad \xi,\omega\in
C_c^\infty\mathrm T^*\mathcal M^{\otimes k}.
\end{equation}
Let us stress that  for non-real $\lambda\in\mathcal D_\alpha$ the
deformed volume form is complex-valued, and the deformed inner
product
$(\xi,\omega)_\lambda=\overline{(\omega,\xi)_{\overline{\lambda}}}$
is not Hermitian. Let $L^2 \mathrm T^*\mathcal M^{\otimes k}$ be the
completion of the set $C_c^\infty \mathrm T^*\mathcal M^{\otimes k}$
with respect to the global inner product $(\cdot,\cdot)\equiv
(\cdot,\cdot)_0$. Observe that Lemma~\ref{l1}  together with the
bounds on $\varrho_\lambda$ implies that the deformed global inner
product $(\cdot,\cdot)_\lambda$ extends  to a bounded
non-degenerate form in $L^2 \mathrm T^*\mathcal M^{\otimes k}$ with
$k=0,1$; i.e. for some $c>0$ and all $\xi,\omega\in L^2 \mathrm
T^*\mathcal M^{\otimes k}$ we have $|(\xi,\omega)_\lambda|^2\leq
c(\xi,\xi)(\omega,\omega)$, and for any nonzero $\xi\in L^2 \mathrm
T^*\mathcal M^{\otimes k}$ there exists  $\omega\in L^2 \mathrm
T^*\mathcal M^{\otimes k}$, such that $(\xi,\omega)_\lambda\neq 0$.
If $\lambda\in\mathcal D_\alpha$ is real, then
$(\cdot,\cdot)_\lambda$ is the global inner product on the
Riemannian manifold $(\mathcal M,\mathsf g_\lambda)$, and
$\sqrt{(\cdot,\cdot)_\lambda}$ is an equivalent norm in $L^2\mathrm
T^*\mathcal M^{\otimes k}$, $k=0,1$.


Let $d: C_c^\infty(\mathcal M)\to C_c^\infty \mathrm T^*\mathcal M$
be  the exterior derivative.  We introduce  the sesquilinear
quadratic form
\begin{equation}\label{q}
\mathsf q_\lambda[u,v]=\bigl(du,d(\overline {\varrho}_\lambda
v)\bigr)_\lambda,\quad u,v\in \mathbf C(\mathsf q_\lambda),\quad
\lambda\in\mathcal D_\alpha,
\end{equation}
where $\mathbf C(\mathsf q_\lambda)$ is a core.
\begin{defn}\label{form core} In the case  $\partial\mathcal M=\varnothing$, and also in the case of the Neumann Laplacian,  we take $\mathbf C(\mathsf q_\lambda)\equiv C_c^\infty(\mathcal M)$. In the case of the Dirichlet Laplacian the core  $\mathbf C(\mathsf q_\lambda)$ consists of the functions $u\in C_c^\infty (\mathcal M)$ with $u\!\upharpoonright_{\partial\mathcal M}=0$.
\end{defn}
The core $\mathbf C(\mathsf q_\lambda)$ is independent of
$\lambda\in\mathcal D_\alpha$ and dense in $L^2(\mathcal M)$.
Moreover, $\mathbf C({^\lambda\!\Delta})\subseteq\mathbf C(\mathsf
q_\lambda)$, cf. Definition~\ref{d1}.
 For all
 $u\in
\mathbf C({^\lambda\!\Delta})$ and $v\in \mathbf C(\mathsf
q_\lambda)$ the Green identity $ (du,d
v)_\lambda=({^{\lambda}\!\Delta u}, v)_\lambda$  can be verified
e.g. in local coordinates. This together with~\eqref{q} and the
definition of the density $\varrho_\lambda$ gives
\begin{equation}\label{stokes}
\mathsf q_\lambda[u,v]=({^{\lambda}\!\Delta u}, v), \quad u\in
\mathbf C({^\lambda\!\Delta}),  v\in \mathbf C(\mathsf q_\lambda).
\end{equation}
Thus to the unbounded operator ${^{\lambda}\!\Delta }$ in
$L^2(\mathcal M)$ with the domain $\mathbf C({^\lambda\!\Delta})$
there corresponds the unbounded quadratic form $\mathsf q_\lambda$
with the domain $\mathbf C(\mathsf q_\lambda)$.

\section{Analytic families of quadratic forms and operators}\label{s4}
In this section we study the unbounded quadratic forms~$\mathsf
q_\lambda$ and the corresponding  unbounded operators
${^\lambda\!\Delta}$ in the Hilbert space $L^2(\mathcal M)$.  We
show that the forms are closable, and their closures define an
analytic family $\mathcal D_\alpha\ni\lambda\mapsto\mathsf
q_\lambda$ in the sense of Kato~\cite{Kato}.  This allows us to see
that the closure of  ${^\lambda\!\Delta}$ is an m-sectorial operator
in $L^2(\mathcal M)$, and the resolvent
$({^\lambda\!\Delta}-\mu)^{-1}$ is an analytic function of
$\lambda$ and $\mu$ on some open subset of $\mathcal
D_\alpha\times\Bbb C$.

We represent the form~\eqref{q} as the sum
\begin{equation}\label{sum}
\mathsf q_\lambda[u,v]=(\varrho_\lambda du,dv)_\lambda+(du,
vd\overline{\varrho}_\lambda)_\lambda
\end{equation}
of two sesquilinear quadratic forms. For the first term in the right
hand side of~\eqref{sum} we prove the following assertion.
\begin{lem}\label{sec form} There exist  $\vartheta<\pi/2$ and $\delta>0$, such that for all $\lambda\in\mathcal D_\alpha$ we have
$$
|\arg (\varrho_\lambda du,du)_\lambda |\leq\vartheta,\
\delta(du,du)_0\leq \Re(\varrho_\lambda du,du)_\lambda\leq
\delta^{-1} (du,du)_0\ \ \forall u\in \mathbf C(\mathsf q_\lambda).
$$
In other words, the form~$(\varrho_\lambda d\cdot,d\cdot)_\lambda$
on the functions in $\mathbf C(\mathsf q_\lambda)$ is sectorial and
relatively bounded, the sector and the bounds are independent of
$\lambda\in\mathcal D_\alpha$.
\end{lem}

\begin{pf} Owing to~\eqref{vol} and~\eqref{form} we have
$$
(\varrho_\lambda d u,d u)_\lambda=\int_{\mathcal M} \mathsf
g_\lambda[ du,du] \operatorname{dvol}_0,\quad u\in
C_c^\infty(\mathcal M),
$$
where $\operatorname{dvol}_0$ is the Riemannian volume form on
$(\mathcal M,\mathsf g)$. Thus the assertion is a direct consequence
of Lemma~\ref{l1}. \qed\end{pf}

On the next step we show that the form $(du,
vd\overline{\varrho}_\lambda)_\lambda$ in the right hand side
of~\eqref{sum}  has an arbitrarily small relative bound with respect
to the  form $(\varrho_\lambda du,dv)_\lambda$ uniformly in
$\lambda$.
\begin{lem}\label{r bound} For any $\epsilon>0$  and all $u\in \mathbf C(\mathsf q_\lambda)$ the estimate
$$
 |(du, ud\overline{\varrho}_\lambda)_\lambda |\leq \epsilon |(\varrho_\lambda du,du)_\lambda |+C\epsilon^{-1}(u,u)
$$
holds, where the constant $C$ is independent of $\epsilon$, $u$, and
$\lambda\in\mathcal D_\alpha$.
\end{lem}
\begin{pf} We have
\begin{equation}\label{s1}
\begin{aligned}
|(du, ud\overline{\varrho}_\lambda)_\lambda | & \leq \int_{\mathcal
M}  | \mathsf g_\lambda [ du , u d\overline{\varrho}_\lambda ]
/{{\varrho}_\lambda}|\operatorname{dvol}_0
\\
& \leq \texttt{C} \Bigl(\int_{\mathcal M}  |\mathsf g_\lambda [du ,
d\overline{\varrho}_\lambda ]|^2 \operatorname{dvol}_0 \Bigr)^{1/2}
\Bigl(\int_{\mathcal M} |u|^2\operatorname{dvol}_0\Bigr)^{1/2}
\\
&\leq \texttt{C}\bigl(\tilde \epsilon \int_{\mathcal M}  |\mathsf
g_\lambda [du , d\overline{\varrho}_\lambda ]|^2
\operatorname{dvol}_0 + \tilde\epsilon^{-1}(u,u)\bigr)
\end{aligned}
\end{equation}
with  arbitrarily small $\tilde\epsilon>0$ and
$\texttt{C}=1/\inf_{p\in\mathcal M} \varrho_\lambda(p)$.
 From Lemma~\ref{l1} it follows that  $|\Im \mathsf g^p_\lambda[\xi,\xi]|\leq (\tan\vartheta)\Re \mathsf g^p_\lambda[\xi,\xi]$, where $\Re \mathsf g^p_\lambda[\cdot,\cdot]$ defines an inner product in the space $\Bbb C\mathrm T^*_p\mathcal M$. This and the Cauchy-Schwarz inequality give
$$
\begin{aligned}
|\mathsf g^p_\lambda [du ,  d\overline{\varrho}_\lambda ]|^2 \leq
(1+\tan\vartheta)^{2}\Re\mathsf g^p_\lambda [du ,  du]\,\Re \mathsf
g^p_\lambda [ d{\varrho}_\lambda , d{\varrho}_\lambda].
\end{aligned}
$$
Evidently $d\varrho_\lambda \equiv 0$ on $\mathcal
M\setminus\{(R,\infty)\times\Omega\}$. Below we establish the
uniform in $\lambda\in\mathcal D_\alpha$ and
$p\in(R,\infty)\times\Omega$ estimate $\Re \mathsf g^p_\lambda[
d{\varrho}_\lambda , d{\varrho}_\lambda] < c$.

 In the local coordinates $(x,y)$ we have
$$
\Re\mathsf g^p_\lambda [ d{\varrho}_\lambda ,
d{\varrho}_\lambda]=\Re(\overline{\xi}\cdot\mathbf
g_\lambda^{-1}(x,y)\xi)\leq \delta^{-1/2}|\xi|^2,
$$
where $\xi=\nabla_{xy}\varrho_\lambda(x,y)$, cf.~\eqref{es}. As $R$
is sufficiently large, from  Lemma~\ref{sss} it follows that
$$
0<c\leq|\det\mathbf g_\lambda(x,y)|\leq 1/c,\quad
|\partial^r_x\partial_y^q\det\mathbf g_\lambda(x,y)|\leq c_{rq},
$$
where $r+|q|= 1$, and the constants $c$ and $c_{rq}$ are independent
of $x>R$, $y\in\kappa_j[\Omega\cap\mathscr U_j]$, and
$\lambda\in\mathcal D_\alpha$. This and~\eqref{mu xy} lead to the
estimate $|\nabla_{xy}\varrho_\lambda(x,y)|^2\leq C$ for all
$\lambda\in\mathcal D_\alpha$, $x>R$, and
$y\in\kappa_j[\Omega\cap\mathscr U_j]$.

 Thus $|\mathsf g^p_\lambda [du ,  d\overline{\varrho}_\lambda ]|^2
\leq c(1+\tan\vartheta)^{2}\Re\mathsf g^p_\lambda [du ,  du]$, and
finally we get
$$
\begin{aligned}
\int_{\mathcal M} |\mathsf g_\lambda [du ,
d\overline{\varrho}_\lambda ]|^2 \operatorname{dvol}_0\leq
c(1+\tan\vartheta)^2\Re \int_{\mathcal M}\mathsf g_\lambda [ du ,
du ] \operatorname{dvol}_0
\\
=c(1+\tan\vartheta)^2\Re (\varrho_\lambda  du,du)_\lambda\leq
c(1+\tan\vartheta)^2|(\varrho_\lambda  du,du)_\lambda|.
\end{aligned}
$$
This together with~\eqref{s1} establishes  the assertion for
$C=\texttt{C}^2 c(1+\tan\vartheta)^2$ and an arbitrarily small
$\epsilon=\texttt{C} c (1+\tan\vartheta)^2\tilde \epsilon$.
\qed\end{pf}

\begin{prop}\label{family of forms}  Introduce the Hilbert space $\mathbf D(\mathsf q_\lambda)$ as the completion of the core  $\mathbf C(\mathsf q_\lambda)$ with respect to the norm $\sqrt{(du,du)_0+\|u\|^2}$. Then the family of unbounded quadratic forms $\mathcal D_\alpha\ni\lambda\mapsto \mathsf q_\lambda$ with the domain $\mathbf D(\mathsf q_\lambda)$ is analytic in the sense of Kato~\cite{Kato}; i.e.  $\mathsf q_\lambda$ is a closed densely defined sectorial form,  $\mathbf D(\mathsf q_\lambda)$ is independent of  $\lambda$, and the function $\mathcal D_\alpha\ni\lambda\mapsto\mathsf q_\lambda[u,u]$ is  analytic for any  $u\in\mathbf D(\mathsf q_\lambda)$. Moreover, the sector of $\mathsf q_\lambda$ is independent of $\lambda$.
\end{prop}
\begin{pf}  It is clear that the nonnegative form $\mathsf q_0$, $\mathsf q_0[u,u]\equiv (du,du)_0$, with the domain $\mathbf D(\mathsf q_\lambda)$ is densely defined and closed. The domain $\mathbf D(\mathsf q_\lambda)$ is independent of $\lambda$, because the core  $\mathbf C(\mathsf q_\lambda)$ and the norm $\sqrt{(du,du)_0+\|u\|^2}$ are independent of $\lambda$, cf. Definition~\ref{form core}.

 As a consequence of the equality~\eqref{sum} and Lemmas~\ref{sec form},~\ref{r bound} we obtain
\begin{equation}\label{i1}
 |\arg(\mathsf q_\lambda [u,u]+\gamma \|u\|^2)|\leq \vartheta,
\end{equation}
\begin{equation}\label{i2}
\delta (du,du)_0\leq \Re \mathsf q_\lambda [u,u]+\gamma\|u\|^2,\quad
\Re \mathsf q_\lambda [u,u]\leq \delta^{-1}
\bigl((du,du)_0+\|u\|^2\bigr)
\end{equation}
with some angle $\vartheta<\pi/2$ and some positive constants
$\delta$ and $\gamma$, which are independent of $u\in \mathbf
C(\mathsf q_\lambda)$ and $\lambda\in\mathcal D_\alpha$.

The estimate~\eqref{i1} shows that $\mathsf q_\lambda$ is a
sectorial form in the space $L^2(\mathcal M)$, and the sector is
independent of $\lambda\in\mathcal D_\alpha$. We recall that {\it
i}) a sequence of functions $\{u_j\}$ is said to be $\mathsf
q_\lambda-$con\-ver\-gent, if $u_j$ is in the domain of $\mathsf
q_\lambda$, $\|u_j-u\|\to 0$ and $\mathsf q_\lambda
[u_j-u_m,u_j-u_m]\to 0$ as $j,m\to \infty$; {\it ii}) the  form
$\mathsf q_\lambda$ is closed, if  every $\mathsf
q_\lambda$-convergent sequence  $\{u_j\}$ has a limit $u$ in the
domain of $\mathsf q_\lambda$, and $\mathsf
q_\lambda[u-u_j,u-u_j]\to 0$. From the inequalities
$$
\Re\mathsf q_\lambda[u,u]+\gamma\|u\|^2\leq |\mathsf
q_\lambda[u,u]+\gamma\|u\|^2|\leq (\cos\vartheta)^{-1}(\Re\mathsf
q_\lambda[u,u]+\gamma\|u\|^2)
$$
together with the estimates~\eqref{i2}, we conclude that the form
$\mathsf q_\lambda$ with the domain $\mathbf D(\mathsf q_\lambda)$
is closed.

Let us show that for any  $u\in\mathbf D(\mathsf q_\lambda)$ the
function $\mathcal D_\alpha\ni\lambda\mapsto\mathsf q_\lambda[u, u]$
is  analytic. Let $\{u_j\}$   be a sequence in $\mathbf C(\mathsf
q_\lambda)$ convergent to $u$ in the norm of  $\mathbf D(\mathsf
q_\lambda)$. Then the analytic functions  $\mathcal
D_\alpha\ni\lambda\mapsto \mathsf q_\lambda[u_j, u_j]$ converge to
the function $\lambda\mapsto \mathsf q_\lambda[u,u]$ uniformly in
$\lambda\in\mathcal D_\alpha$  as $j\to+\infty$.  Indeed, as a
consequence of the  estimates~\eqref{i1} and~\eqref{i2} we get
  \begin{equation*}
  \begin{aligned}
 |&\mathsf q_\lambda  [ u_j,  u_j]-  \mathsf q_\lambda[ u,  u]|^2\leq  4|\mathsf q_\lambda[u_j-u,u_j]+\gamma(u_j-u,u_j)_0|^2
 \\
  &+  4|\mathsf q_\lambda[u,u-u_j]+\gamma(u,u-u_j)_0|^2+2\gamma^2(\|u_j\|^2-\|u\|^2)^2
  \\
    &\leq  4(1+\tan\vartheta)^{2}\bigl( \delta^{-1}((du_j,du_j)_0+(du,du)_0)+(\gamma+\delta^{-1})(\|u_j\|^2+\|u\|^2)\bigr)
  \\
  &\ \times\bigl(\delta^{-1}(du_j-du,du_j-du)_0+(\gamma+\delta^{-1})\|u_j-u\|^2\bigr)+2\gamma^2\bigl(\|u\|^2-\|u_j\|^2\bigr)^2,
  \end{aligned}
  \end{equation*}
  where the right hand side is  independent of $\lambda$, and tends to zero  as $j\to\infty$. Hence the limit $\mathcal D_\alpha\ni\lambda\mapsto\mathsf q_\lambda[u, u]$  is an analytic function.
\qed\end{pf}
\begin{thm}\label{AF} Let the assumptions of Theorem~\ref{T1} be fulfilled. Then the following assertions are valid.
\begin{itemize}
\item[1.] The operator ${^\lambda\!\Delta}$ in $L^2(\mathcal M)$, initially defined on the core $\mathbf C({^\lambda\!\Delta})$, admits a closure. The domain of the closure will be denoted by $\mathbf D({^\lambda\!\Delta})$.

\item[2.]  The operator ${^\lambda\!\Delta}$  with the domain $\mathbf D({^\lambda\!\Delta})$ is m-sectorial, and its sector is independent of $\lambda\in\mathcal D_\alpha$.
(Here m-sectorial means that the numerical range $\{z\in\Bbb C:
z=({^\lambda\!\Delta}u,u), u\in\mathbf D({^\lambda\!\Delta}) \}$ and
the spectrum $\sigma({^\lambda\!\Delta})$ of the operator
${^\lambda\!\Delta}$ both lie in some sector $\{z\in\Bbb C:|\arg
(z-\gamma)|\leq \vartheta\}$ of angle $2\vartheta<\pi$.) In
particular, the Laplacian $\Delta\equiv{^0\!\Delta}$ is a
nonnegative selfadjoint operator.

\item[3.]  The resolvent $\Gamma\ni (\lambda,\mu)\mapsto ({^\lambda\!\Delta}-\mu)^{-1}$ is an analytic
    function of two variables on the open set $\Gamma=\bigl\{(\lambda,\mu): \lambda\in\mathcal D_\alpha, \mu\in\Bbb C\setminus\sigma( {^\lambda\!\Delta})\bigr\}$.
\end{itemize}
\end{thm}
\begin{pf}
By Proposition~\ref{family of forms} the form $\mathsf q_\lambda$ is
densely defined, sectorial, and closed. The equality~\eqref{stokes}
extends by continuity to all $v\in\mathbf D(\mathsf q_\lambda)$ and
$u\in\mathbf C({^\lambda\!\Delta})$. This implies that  the deformed
Laplacian ${^\lambda\!\Delta}$, initially defined on the dense in
$\mathbf D(\mathsf q_\lambda)$ core $\mathbf C({^\lambda\!\Delta})$,
is closable. The closure is an m-sectorial operator. Indeed, the set
of all densely defined closed sectorial sesquilinear forms and the
set of all m-sectorial operators are in  one-to-one correspondence
e.g.~\cite[Chapter VI.2.1]{Kato}. Namely, to every form $\mathsf
q_\lambda$ there corresponds a unique m-sectorial operator
${^\lambda\!\Delta}$ in $L^2(\mathcal M)$, such that its domain
$\mathbf D ({^\lambda\!\Delta})$ is dense in $\mathbf D(\mathsf
q_\lambda)$, and $\mathsf q_\lambda[u,v]=({^\lambda\!\Delta}u,v)$
for all $u\in \mathbf D({^\lambda\!\Delta})$ and  $v\in\mathbf
D(\mathsf q_\lambda)$. Moreover,  the sector of ${^\lambda\!\Delta}$
coincides with the sector of the form $\mathsf q_\lambda$. In
particular, to the symmetric nonnegative form $\mathsf q_0$ there
corresponds a nonnegative selfadjoint operator
${^0\!\Delta}\equiv\Delta$. The assertions {\it 1} and {\it 2} are
proven.

 By Proposition~\ref{family of forms} the  family of forms $\mathcal D_\alpha\ni\lambda\mapsto \mathsf q_\lambda$ is analytic in the sense of Kato. By definition this means that the family of m-sectorial operators $\mathcal D_\alpha\ni\lambda\mapsto {^\lambda\!\Delta}$ is an analytic family of type $(\mathbf B)$. 
 As is known,  any analytic family of type $(\mathbf B)$ is also an analytic family of operators in the sense of Kato e.g.~\cite{Simon Reed iv, Kato}. Now a standard argument justifies the assertion~{\it 3}, e.g. \cite[Theorem XII.7]{Simon Reed iv}.
\qed\end{pf}

\section{Localization of the essential spectrum}\label{s5}


 Consider the domain $\mathbf D({^\lambda\!\Delta})$  of the  m-sectorial operator ${^\lambda\!\Delta}$   as a Hilbert
space  with  the norm
$\sqrt{\|\cdot\|^2+\|{^\lambda\!\Delta}\cdot\|^2}$. We say that
$\mu$ is a point of the essential spectrum
$\sigma_{ess}({^\lambda\!\Delta})$, if the bounded operator $
{^\lambda\!\Delta}-\mu :\mathbf D({^\lambda\!\Delta})\to
L^2(\mathcal M) $ is not  Fredholm. Recall that  a bounded linear
operator is said to be  Fredholm (or, equivalently, possesses the
Fredholm property), if  its kernel  and cokernel
 are finite-dimensional, and the
range  is closed. In this section we prove the following theorem.
\begin{thm}\label{ess}
Let the assumptions of Theorem~\ref{T1} be fulfilled,and let
$\lambda\in \mathcal D_\alpha$ be  fixed.   Then the  continuous
operator ${^\lambda\!\Delta}-\mu:\mathbf D({^\lambda\!\Delta})\to
L^2(\mathcal M)$ is Fredholm, if and only if $\mu\in\Bbb C$ does not
meet the condition~\eqref{eq9}.
\end{thm}
The proof of Theorem~\ref{ess} is essentially based on methods of
the theory of elliptic non-homogeneous boundary value
problems~\cite{MP2,KozlovMaz`ya,KozlovMazyaRossmann}. This theory
provides us with necessary and sufficient conditions for the
Fredholm property of  operators of non-homogeneous elliptic boundary
value problems on $\mathcal M$ under an assumption on  stabilization
of their coefficients  at infinity.  In fact, the needed
stabilization of the coefficients of the operator $^\lambda\!\Delta$
(and also of the operator $^\lambda\!\partial_\nu$, if
$\partial\mathcal M\neq\varnothing$) is guaranteed by the
relation~\eqref{4} in Lemma~\ref{sss} and the local
representations~\eqref{LR}. Once the ellipticity of the deformed
Laplacian is established,  Theorem~\ref{ess} can be obtained by
known methods of the mentioned elliptic theory, except for necessity
of the condition on $\mu$ in the case $\partial \mathcal M\neq
\varnothing$.

In the next two lemmas we show that the deformed Laplacian is an
elliptic operator on $\mathcal M$. Then we prove Theorem~\ref{ess}.
\begin{lem}\label{seLem}
Let the assumptions of Theorem~\ref{T1} be fulfilled.  Then for all
$\lambda\in\mathcal D_\alpha$ the deformed Laplacian
${^\lambda\!\Delta}$ is a strongly elliptic differential operator on
the manifold $\mathcal M$.
\end{lem}
\begin{pf}
 Since
$\Delta$ is a strongly elliptic operator, and
$\Delta\equiv{^\lambda\!\Delta}$ outside of the subset
$(R,\infty)\times\Omega$ of the manifold $\mathcal M$, we only need
to check that the deformed Laplacian ${^\lambda\!\Delta}$ is
strongly elliptic on $(R,\infty)\times\Omega$, provided that  $R>0$
is sufficiently large. (We recall that ${^\lambda\!\Delta}$ depends
on $R$, however we do not indicate this for brevity of notations.)

We will rely on the representation~\eqref{LR} of the operator
$^\lambda\!\Delta$ in the local coordinates. Consider the principal
symbol $\xi \cdot\mathbf g^{-1}_\lambda(x,y)\xi$ of
${^\lambda\!\Delta}$ on $\Pi$, where $\xi\in\Bbb R^{n+1}$ and
$(x,y)\in(R,\infty)\times\kappa_j[\mathscr U_j\cap\Omega]$. The
operator ${^\lambda\!\Delta}$ is strongly elliptic due to the
estimate~\eqref{es} established in the proof of Lemma~\ref{l1}.
\qed\end{pf}

In the case of a manifold $\mathcal M$ without boundary the deformed
Laplacian is an elliptic operator by Lemma~\ref{seLem}.
  In  the case $\partial\mathcal M\neq\varnothing$ we also need to verify the validity of the Shapiro-Lopatinski\v{\i} condition on $\partial\mathcal M$. 
As is well known, the Shapiro-Lopatinski\v{\i} condition is always
satisfied for a strongly elliptic operator and the Dirichlet
boundary condition, e.g.~\cite{Lions Magenes}. Therefore the
deformed Dirichlet Laplacian is an elliptic operator on $\mathcal
M$. In the following lemma we check the validity of the
Shapiro-Lopatinski\v{\i} condition  for the deformed Neumann
Laplacian.

\begin{lem}\label{S-L}
Let the assumptions of Theorem~\ref{T1} be fulfilled, and $\partial
\mathcal M\neq\varnothing$.  Then for all $\lambda\in\mathcal
D_\alpha$   the strongly elliptic operator ${^\lambda\!\Delta}$ and
the operator ${^\lambda\!\partial_\nu}$ of the deformed Neumann
boundary condition  satisfy the Shapiro-Lopatinski\v{\i} condition
on $\partial \mathcal M$. In other words, the deformed Neumann
Laplacian is an elliptic operator on $\mathcal M$.
\end{lem}
\begin{pf}
The complex scaling does not deform the  Laplacian and the operator
of the Neumann boundary condition outside of
$(R,\infty)\times\Omega\subset\mathcal M$. As is well known,  the
Neumann Laplacian satisfies the Shapiro-Lopatinski\v{\i} condition,
e.g.~\cite{Taylor}. Thus we only need to prove that the
Shapiro-Lopatinski\v{\i} condition is valid for the strongly
elliptic operator ${^\lambda\!\Delta}$ on $(R,\infty)\times\Omega$
with the operator ${^\lambda\!\partial_\nu}$ of the deformed Neumann
boundary conditions on $(R,\infty)\times\partial\Omega$.  We recall
that the operators ${^\lambda\!\Delta}$ and
${^\lambda\!\partial_\nu}$ both depend on $R$, however we do not
indicate this for brevity of notations.

The scheme of the proof is as follows. We first consider an
auxiliary boundary value problem on $(R,\infty)\times\Omega$ for a
strongly elliptic operator, and show that this problem satisfies the
Shapiro-Lopatinski\v{\i} condition. As is known, this condition
implies that the corresponding parameter-dependent problem on the
half-axis $\Bbb R_+$ is uniquely solvable in the scale of Sobolev
spaces. Then we consider the parameter-dependent problem on $\Bbb
R_+$ that corresponds to the deformed Neumann Laplacian. We prove
that this homogeneous problem has no nontrivial solutions in the
Sobolev space $H^2(\Bbb R_+)$, and therefore the deformed Neumann
Laplacian satisfies the Shapiro-Lopatinski\v{\i} condition. Here we
rely on the fact that for large $R$ the operators of these two
problems on $\Bbb R_+$ are close to each other in the operator norm.
This is a consequence of the estimate~\eqref{3} in Lemma~\ref{sss}.

Recall that  $\Delta_\Omega$ is the Laplacian on the compact
manifold $(\Omega,\mathfrak h)$. Consider the auxiliary operator
$\Delta_\Omega-\bigl((1+\lambda \mathsf
s'_R(x))^{-1}\partial_x\bigr)^2$  on $(R,\infty)\times\Omega$.
Observe that its principal symbol is
$\xi\cdot\overset{\infty}{\mathbf g}\vphantom{\mathbf
g}^{-1}_\lambda(x,y)\xi$, where $\xi\in \Bbb R^{n+1}$, $(x,y)\in
(R,\infty)\times \kappa_j[\mathscr U_j\cap\Omega]$, and
$\overset{\infty}{\mathbf g}\vphantom{\mathbf g}_\lambda(x,y)$ is
the matrix~\eqref{h lambda}. Therefore the operator  is strongly
elliptic due to the inequalities~\eqref{eee-1}.

Let  $\partial_\eta$ be the operator of the Neumann boundary
conditions on $(R,\infty)\times\partial \Omega$ taken with respect
to the product metric $\overset{\infty}{\mathsf g}=dx \otimes
dx+\mathfrak h$.  Let us  check the validity of the
Shapiro-Lopatinski\v{\i} condition on $(R,\infty)\times\partial
\Omega$ for the pair
\begin{equation}\label{pair}
\bigl\{\Delta_\Omega-\bigl((1+\lambda \mathsf
s'_R(x))^{-1}\partial_x\bigr)^2,
\partial_\eta\bigr\}.
\end{equation}

Let $\{\mathscr U_j\}$ be a sufficiently fine open cover of
$\Omega$. In every neighborhood $\mathscr U_j$ with $\mathscr
U_j\cap\partial\Omega\neq\varnothing$ we pick boundary normal
coordinates $y=(y',y_n)$ on $(\Omega,\mathfrak h)$,  such that
$\partial_{y_n}$ coincides with the unit inward normal derivative
given by the metric $\mathfrak h$. Then $(x,y)$ are local boundary
normal coordinates  on the semi-cylinder
$(\Pi,\overset{\infty}{\mathsf g})$. In this coordinates we have the
product metric representation
$$
\overset{\infty}{\mathsf g}(y)=dx \otimes
dx+\sum_{m,\ell=1}^{n-1}\mathbf h_{m\ell}(y)\,dy_m\otimes dy_\ell +d
y_n\otimes d y_n.
$$

 The
principal part of the Laplacian $\Delta_\Omega$ at the points
$(y',0)$ of $\mathscr U_j\cap\partial\Omega$ has the form
$-Q(y',\partial_{y'})-\partial^2_{y_n}$. Here $Q( y', \xi')$ is a
positive definite quadratic form in $\xi'\in\Bbb R^{n-1}$, whose
coefficients are functions of $y'$. Let $\xi=(\xi_0,\xi')\in\Bbb
R^n$.
 The  Shapiro-Lopatinski\v{\i} condition for the pair~\eqref{pair}  is satisfied at a point $(x, y',0)$, if for all $\xi\in\Bbb R^n\setminus\{0\}$  the parameter-dependent problem on $\Bbb R_+$
\begin{equation}\label{ap}
\bigl((1+\lambda \mathsf s'_R(x))^{-2}\xi_0^2+Q(y',
\xi')-\partial_{y_n}^2\bigr)\mathsf u(y_n)=0\text{ for } y_n\in\Bbb
R_+,\ \  \partial_{y_n} \mathsf u(0)=0
\end{equation}
has no nontrivial solutions, which are exponentially decaying as
$y_n\to+\infty$.   For $\xi\neq 0$, any bounded solution
of~\eqref{ap} must be a multiple of
$$
e^{-y_n\sqrt{(1+\lambda \mathsf s'_R(x))^{-2}\xi_0^2+Q(y',\xi')}}.
$$
Due to the inequalities $0\leq \mathsf s'_R(x)\leq 1$ and
$|\lambda|<\sin\alpha<2^{-1/2}$ we have  $\Re (1+\lambda \mathsf
s'_R(x))^{-2}>0$, and therefore
$$
\partial_{y_n} e^{-y_n\sqrt{(1+\lambda \mathsf s'_R(x))^{-2}\xi_0^2+Q( y', \xi')}}\upharpoonright_{y_n=0}=-\sqrt{(1+\lambda \mathsf s'_R(x))^{-2}\xi_0^2+Q(y',\xi')}\neq 0
$$
for all $\xi\in\Bbb R^n\setminus\{0\}$. Thus the pair~\eqref{pair}
satisfies the Shapiro-Lopatinski\v{\i} condition on
$(R,\infty)\times\partial \Omega$.

On the other hand, it is well known \cite{KozlovMazyaRossmann,
Taylor, Lions Magenes} that the pair~\eqref{pair} satisfies the
Shapiro-Lopatinski\v{\i} condition at  $(x,y',0)$, if and only if
for all $\xi\in\Bbb R^n\setminus\{0\}$ the mapping
$$
H^2(\Bbb R_+)\ni \mathsf u\mapsto \bigl\{\bigl((1+\lambda \mathsf
s'_R(x))^{-2}\xi_0^2+Q( y', \xi')-\partial_{y_n}^2\bigr)\mathsf
u,\partial_{y_n}\mathsf u (0)\bigr\}\in L^2(\Bbb R^+)\times \Bbb C
$$
realizes an isomorphism; here the space $L^2(\Bbb R_+)$ and the
Sobolev space $H^2(\Bbb R_+)$ are endowed with the usual norms
 $$
 \|\mathsf f\|_{
L^2(\Bbb R_+)}=\Bigl(\int_0^\infty|\mathsf f(y_n)|^2\,d
y_n\Bigr)^{1/2}, \quad
 \|\mathsf u\|_{ H^2(\Bbb R_+)}=\Bigl(\sum_{j\leq 2}\|\partial_{y_n}^j \mathsf u\|_{ L^2(\Bbb R_+)}^2\Bigr)^{1/2}.
 $$
Hence for all $\xi\in\Bbb R^n\setminus\{0\}$ and $\mathsf u\in
H^2(\Bbb R_+)$ the estimate
\begin{equation}\label{est}
\begin{aligned}
\|\mathsf u\|_{ H^2(\Bbb R_+)}\leq C\bigl(\bigl\|\bigl((1+\lambda
\mathsf s'_R(x))^{-2}\xi_0^2+Q(
y',\xi')-\partial_{y_n}^2\bigr)\mathsf u\bigr\|_{ L^2(\Bbb
R_+)}+|\partial_{y_n}\mathsf u(0)|\bigr)
\end{aligned}
\end{equation}
holds, where the constant $C$ may depend on $\xi$, $x$, $y'$,
$\lambda$, and $R$, but not on $\mathsf u$. Our next aim is to show
that  there exists a universal constant $C$, such that the
estimate~\eqref{est} remains valid for all $\xi\in{\mathsf
S}^n=\{\xi\in\Bbb R^n: |\xi|= 1\}$, $x>R$, $y'$,
$\lambda\in\mathcal D_\alpha$, and $R\geq 0$.

Observe first that the constant $C$ in~\eqref{est} is independent of
$R$, because the estimate~\eqref{est} with  $R$ replaced by $\tilde
R$ can be obtained from~\eqref{est} by the change of  variables
$x\mapsto x+R-\tilde R$. Without loss of generality we set $R=0$.

The function $\mathsf s'\equiv \mathsf s'_0$, and therefore the
estimate~\eqref{est}, depend on $x$ only on a compact  subset
$\mathsf K$ of $\Bbb R_+$,  cf.~\eqref{ab}. We can cover the compact
set
 $
 \mathsf K\times\overline{\mathcal D_\alpha}\times {\mathsf S}^n
 $
by a finite number of sufficiently small neighborhoods. As before we
justify the estimate~\eqref{est} for a fixed point $(x,\lambda, \xi,
y')$. As $(\tilde x,\tilde\lambda, \tilde\xi, \tilde y')$ varies in
a sufficiently small neighborhood of $(x,\lambda, \xi, y')$,  the
estimate
\begin{equation}\label{est2}
\begin{aligned}
\bigl\|\bigl((1+\tilde \lambda \mathsf s'(\tilde
x))^{-2}\tilde\xi_0^2+Q(\tilde y', &\tilde\xi')\bigr)\mathsf u
\\
-\bigl((1+\lambda \mathsf s'(x))^{-2}\xi_0^2+ &Q(
y',\xi')\bigr)\mathsf u\bigr\|_{ L^2(\Bbb R_+)}\leq
\epsilon\|\mathsf u\|_{ H^2(\Bbb R^+)}
\end{aligned}
\end{equation}
remains valid  for some $\epsilon<1/C$. Then~\eqref{est2} together
with~\eqref{est} implies
\begin{equation}\label{est3}
\begin{aligned}
\|\mathsf u\|_{ H^2(\Bbb R_+)}\leq
(1/C&-\epsilon)^{-1}\bigl(\bigl\|\bigl((1+\lambda \mathsf s'(\tilde
x))^{-2}\tilde\xi_0^2
\\
&+Q( \tilde y',\tilde \xi')-\partial_{y_n}^2\bigr)\mathsf u
\bigr\|_{ L^2(\Bbb R_+)}+|\partial_{y_n}\mathsf u(0)|\bigr).
\end{aligned}
\end{equation}
We have proved that there exists $C<\infty$, such that the
estimate~\eqref{est} is valid for all $\xi\in{\mathsf S}^n$, $x>R$,
$(y',0)\in\kappa_j[\partial\Omega\cap\mathscr U_j] $,
$\lambda\in\mathcal D_\alpha$, and $R\geq 0$.

Let $\mathbf g_\lambda(x,y)$ be the matrix~\eqref{metric}. We take
the principal part of the deformed Laplacian $^\lambda\!\Delta$
written in the local coordinates, cf.~\eqref{LR}.  Then we freeze
the coefficients of the principal part at a point of the boundary,
and replace $\partial_x$ by $-i\xi_0$ and $\partial_{y'}$ by
$-i\xi'$. As a result we get the parameter-dependent operator
$$
(\xi,i\partial_{y_n})\mathbf g_\lambda^{-1}
(x,y',0)(\xi,i\partial_{y_n})^\intercal, \quad y_n\in\Bbb R_+.
$$
The deformed Neumann boundary condition $^\lambda\!\partial_\nu u=0$
can be written in the local coordinates as follows
$$
\bigl(0,\dots,0,1/\sqrt{\mathbf
g^{-1}_{\lambda,nn}(x,y',0)}\bigr)\mathbf
g^{-1}_\lambda(x,y',0)\nabla_{xy}u(x,y',0)=0,
$$
cf.~\eqref{LR}. The deformed Neumann Laplacian satisfies the
Shapiro-Lopatinski\v{\i} condition at  $(x,y',0)$, if for all
$\xi\in\Bbb R^n\setminus\{0\}$  the  problem
\begin{equation}\label{pr}
\begin{aligned}
&(\xi,i\partial_{y_n})\mathbf g_\lambda^{-1} (x,y',0)(\xi,i\partial_{y_n})^\intercal \mathsf u(y_n)=0\text{ for } y_n\in\Bbb R_+,\\
& \bigl(0,\dots,0,1/\sqrt{\mathbf
g^{-1}_{\lambda,nn}(x,y',0)}\bigr)\mathbf
g^{-1}_\lambda(x,y',0)(\xi,i\partial_{y_n})^\intercal \mathsf u(0)=0
\end{aligned}
\end{equation}
has no nontrivial solutions, which are exponentially decaying as
$y_n\to+\infty$.

In the boundary normal coordinates we have
$$
\begin{aligned}
& (1+\lambda \mathsf s'_R(x))^{-2}\xi_0^2+Q(
y',\xi')-\partial_{y_n}^2=(\xi,i\partial_{y_n})\overset{\infty}{\mathbf
g}\vphantom{\mathbf g}_\lambda ^{-1}
(x,y',0)(\xi,i\partial_{y_n})^\intercal,
\\
& i\partial_{y_n }=\bigl(0,\dots,0,1/\sqrt{\overset{\infty}{\mathbf
g}\vphantom{\mathbf
g}^{-1}_{\lambda,nn}(x,y',0)}\bigr)\overset{\infty}{\mathbf
g}\vphantom{\mathbf
g}^{-1}_\lambda(x,y',0)(\xi,i\partial_{y_n})^\intercal,
\end{aligned}
$$
where $\overset{\infty}{\mathbf g}\vphantom{\mathbf g}_\lambda(x,y)$
is the matrix~\eqref{h lambda}.

Recall that the matrices $\mathbf g_\lambda(x,y)$ and
$\overset{\infty}{\mathbf g}\vphantom{\mathbf g}_\lambda(x,y)$ both
depend on $R$. Moreover, by virtue of Lemma~\ref{sss}, $\mathbf
g^{-1}_\lambda(x,y)$ tends to $\overset{\infty}{\mathbf
g}\vphantom{\mathbf g}^{-1}_\lambda(x,y)$  uniformly in $x>R$, $y\in
\kappa_j[\Omega\cap\mathscr U_j]$, and $\lambda\in\mathcal D_\alpha$
as $R\to+\infty$. As a consequence we have the uniform in
$\xi\in{\mathsf S}^n$, $x>R$, $y'$, and $\lambda\in\mathcal
D_\alpha$ estimates
$$
\begin{aligned}
\|(\xi,i\partial_{y_n})\bigl(\mathbf
g_\lambda^{-1}(x,y',0)-\overset{\infty}{\mathbf g}\vphantom{\mathbf
g}_\lambda^{-1}(x,y',0) \bigr)(\xi,i\partial_{y_n})^\intercal &
\mathsf u\|_{ L^2(  \Bbb R_+)}\leq c(R)\|\mathsf u\|_{ H^2(\Bbb
R_+)},
\\
\bigl|\bigl(0,\dots,0,1/\sqrt{\mathbf
g^{-1}_{\lambda,nn}(x,y',0)}\bigr)\mathbf
g^{-1}_\lambda(x,y',0)(\xi,i\partial_{y_n})^\intercal   & \mathsf
u(0)-i\partial_{y_n} \mathsf u(0)\bigr|\\&\leq c(R) \|\mathsf u\|_{
H^2(\Bbb R_+)},
\end{aligned}
$$
where $c(R)\to 0$ as $R\to +\infty$. These estimates together
with~\eqref{est} imply that
\begin{equation}\label{F-1}
\begin{aligned}
\|\mathsf u\|_{ H^2(\Bbb R_+)}\leq \mathrm C(R)
\bigl(\|(\xi,i\partial_{y_n})\mathbf g_\lambda^{-1}
(x,y',0)(\xi,i\partial_{y_n})^\intercal \mathsf u\|_{ L^2(\Bbb R_+)}
\\
+\bigl|\bigl(0,\dots,0,1/\sqrt{\mathbf
g^{-1}_{\lambda,nn}(x,y',0)}\bigr)\mathbf
g^{-1}_\lambda(x,y',0)(\xi,i\partial_{y_n})^\intercal \mathsf
u(0)\bigr|\bigr),
\end{aligned}
\end{equation}
where the constant $\mathrm C(R) =(1/C-c(R))^{-1}$ is positive (for
all sufficiently large  $R$) and independent of  $\xi\in{\mathsf
S}^n$, $x>R$, $y'$, and $\lambda\in\mathcal D_\alpha$.

Substituting $\mathsf u(y_n)=\mathsf v(|\xi|y_n)$ into the
problem~\eqref{pr}, we get
\begin{equation}\label{pr2}
\begin{aligned}
&(|\xi|^{-1}\xi,i\partial_{y_n})\mathbf g_\lambda^{-1} (x,y',0)(|\xi|^{-1}\xi,i\partial_{y_n})^\intercal \mathsf v(y_n)=0\text{ for } y_n\in\Bbb R_+,\\
& \bigl(0,\dots,0,1/\sqrt{\mathbf
g^{-1}_{\lambda,nn}(x,y',0)}\bigr)\mathbf
g^{-1}_\lambda(x,y',0)(|\xi|^{-1}\xi,i\partial_{y_n})^\intercal
\mathsf v(0)=0.
\end{aligned}
\end{equation}
Consequently, the problem~\eqref{pr} has an exponentially decaying
solution $\mathsf u$ for some $\xi\in\Bbb R^n\setminus\{0\}$, if and
only if there exists an exponentially decaying solution $\mathsf v$
of the problem~\eqref{pr2}, where $|\xi|^{-1}\xi\in{\mathsf S}^n$.
In the estimate~\eqref{F-1} we replace $\mathsf u$ by $\mathsf v$
and $\xi$ by $|\xi|^{-1}\xi$, and conclude that the
problem~\eqref{pr2} with $|\xi|^{-1}\xi\in{\mathsf S}^n$ has no
nontrivial solutions $\mathsf v\in H^2(\Bbb R_+)$. Therefore for any
$\xi\in\Bbb R^n\setminus\{0\}$  the problem~\eqref{pr} has no
exponentially decaying solutions. This proves that there exists a
sufficiently large $R>0$, such that the deformed Neumann Laplacian
is an elliptic operator on $\mathcal M$ for all values of the
scaling parameter $\lambda$ in the disk $\mathcal D_\alpha$.
\qed\end{pf}

\begin{pf*}{Proof of Theorem~\ref{ess}.} We prove the theorem for the case of the deformed Neumann Laplacian $^\lambda\!\Delta$ on $\mathcal M$. In the case $\partial\mathcal M=\varnothing$, as well as in the case of the deformed Dirichlet Laplacian, the proof is similar, and in fact simpler, cf.~\cite{Kalvin}. We leave it to the reader.

We  will rely on the following  lemma  due to Peetre, see
e.g.~\cite[Lemma~5.1]{Lions Magenes},
\cite[Lemma~3.4.1]{KozlovMazyaRossmann} or~\cite{Peetre}:
\begin{itemize}
\item[]{\it Let $\mathcal X,\mathcal Y$ and $\mathcal Z$ be Banach spaces, where $\mathcal X$ is compactly embedded into $\mathcal Z$. Furthermore, let $\mathcal L$ be a linear continuous operator from $\mathcal X$ to $\mathcal Y$. Then the next two assertions are equivalent: (i) the range of $\mathcal L$ is closed in $\mathcal Y$ and $\dim \ker \mathcal L<\infty$, (ii) there exists a constant $C$, such that
    \begin{equation}\label{coercive}
    \|u\|_{\mathcal X}\leq C(\|\mathcal L u\|_{\mathcal Y}+\|u\|_{\mathcal Z})\quad \forall u\in \mathcal X.
    \end{equation}}
\end{itemize}

{\it Sufficiency.} Here we assume that the spectral parameter $\mu$
does not meet the condition~\eqref{eq9}, and establish an estimate
of type~\eqref{coercive} for the deformed Neumann Laplacian.

 We first  show that the operator $^\lambda\!\Delta$ on $\mathcal M$ and  the operator $^\lambda\!\partial_\nu$ of the deformed Neumann boundary condition on $\partial\mathcal M$ stabilize at infinity in a certain sense.

  Consider the differential operator $\Delta_\Omega-(1+\lambda)^{-2}\partial_x^2$ in the infinite cylinder $\Bbb R\times\Omega$. By $\partial_\eta$ we denote the operator of the Neumann boundary condition on $\Bbb R\times\partial\Omega$, taken with respect to the product metric   $dx\otimes dx+\mathfrak h$ on $\Bbb R\times\Omega$.  Let $T>0$ be so large that $\mathsf s'_R(x)=1$ for all $x\geq T$ (here $\mathsf s_R(x)=\mathsf s(x-R)$ with a function $\mathsf s$ obeying the conditions~\eqref{ab} and a sufficiently large fixed parameter $R$). Then for all $x\geq T$ we can write the differences
 ${^\lambda\!\Delta}-(
\Delta_\Omega-(1 + \lambda)^{-2}\partial_x^2)$ and
${^\lambda\!\partial_\nu}-\partial_\eta$  in the local coordinates
$(x,y)$ on $\Pi$ as follows:
\begin{equation}\label{olc}
\begin{aligned}
\frac 1 {\sqrt{\det \mathbf g_\lambda}}\nabla_{xy}\cdot\sqrt{\det
\mathbf g_\lambda}\mathbf g_\lambda^{-1}\nabla_{xy}-\frac 1
{\sqrt{\det \overset{\infty}{\mathbf g}\vphantom{\mathbf
g}_\lambda}}\nabla_{xy}\cdot\sqrt{\det \overset{\infty}{\mathbf
g}\vphantom{\mathbf g}_\lambda}\overset{\infty}{\mathbf
g}\vphantom{\mathbf g}_\lambda^{-1}\nabla_{xy}
,\\
\Bigl(\bigl(0,\dots,0,1/\sqrt{\overset{\infty}{\mathbf
g}\vphantom{\mathbf
g}^{-1}_{\lambda,nn}}\bigr)\overset{\infty}{\mathbf
g}\vphantom{\mathbf g}^{-1}_\lambda-\bigl(0,\dots,0,1/\sqrt{\mathbf
g^{-1}_{\lambda,nn}}\bigr)\mathbf
g^{-1}_\lambda\Bigr)\upharpoonright_{y_n=0}\nabla_{xy},
\end{aligned}
\end{equation}
where $\mathbf g_\lambda$ is the matrix~\eqref{metric}, and
$\overset{\infty}{\mathbf g}\vphantom{\mathbf g}_\lambda$ is the
matrix~\eqref{h lambda}. Due to  the property~\eqref{4} of  $\mathbf
g_\lambda^{-1}$,  the coefficients of the operators~\eqref{olc}
uniformly tend to zero  as $x\to +\infty$. In this sense
$^\lambda\!\Delta$ and $^\lambda\!\partial_\nu$ stabilize at
infinity to the operators
$\Delta_\Omega-(1+\lambda)^{-2}\partial_x^2$ and $\partial_\eta$,
whose coefficients are independent of $x$.

 Introduce the Sobolev space $H^\ell(\Bbb R\times\Omega)$ of functions on the infinite cylinder $\Bbb R\times \Omega$ as the completion of the set $C_c^\infty(\Bbb R\times\Omega)$ with respect to the norm
$$
\|u\|_{H^\ell(\Bbb R\times\Omega)}=\Bigl (\int_{\Bbb R}\sum_{r\leq
\ell}\|\partial^r_x u\|^2_{ H^{\ell-r}(\Omega)}\,dx\Bigr)^{1/2},
$$
where $H^\ell(\Omega)$ is the Sobolev space of functions on the
compact manifold $\Omega$.
  Here and elsewhere the norms in the Sobolev spaces on smooth compact manifolds are defined in local coordinates with the help of a finite partition of unity; e.g. $H^\ell(\Omega)$ is the completion of the set $C_c^\infty(\Omega)$ with respect to the norm
$$
\|u\|_{H^\ell(\Omega)}=\Bigl(\sum_j
\int_{\kappa_j[\Omega\cap\mathscr U_j]}\sum_{|q|\leq
\ell}|\partial_y^q (\phi_j u)(y)|^2\,dy\Bigr)^{1/2},
$$
where $\{\mathscr U_j,\kappa_j\}$ is the atlas on $\Omega$, and
$\{\phi_j\}$ is a partition of unity subordinated to the cover
$\{\mathscr U_j\}$. Although the norm in $H^\ell(\Omega)$ depends on
the atlas and the partition of unity, the norms corresponding to
different atlases and partitions of unity are equivalent.

Let $H^{1/2}(\Bbb R\times\partial\Omega)$ be the space of traces
$u\!\upharpoonright_{\Bbb R\times\partial\Omega}$ of the functions
$u\in H^1(\Bbb R\times\Omega)$. By applying the Fourier transform
$\mathcal F_{x\mapsto \xi}$ we pass from the continuous operator
\begin{equation}\label{opo}
\{\Delta_\Omega-(1+\lambda)^{-2}\partial_x^2-\mu,\partial_\eta\}:
H^2(\Bbb R\times\Omega)\to H^0(\Bbb R\times\Omega)\times
H^{1/2}(\Bbb R\times\partial\Omega)
\end{equation}
of the Neumann boundary value problem in the infinite cylinder $\Bbb
R\times\Omega$ to the operator
$\{\Delta_\Omega+(1+\lambda)^{-2}\xi^2-\mu,\partial_\eta\}$ of the
Neumann boundary value problem  on  $(\Omega,\mathfrak h)$. Assume
that $\mu$ does not meet the condition~\eqref{eq9} or, equivalently,
that  for any $\xi\in\Bbb R$ the number $\mu-(1+\lambda)^{-2}\xi^2$
is not an eigenvalue $\nu_j$ of the Neumann Laplacian on
$(\Omega,\mathfrak h)$.  Then a known argument, see
e.g.~\cite[Theorem 4.1]{MP2} or \cite[Theorem
5.2.2]{KozlovMazyaRossmann} or \cite[Theorem~2.4.1]{KozlovMaz`ya},
shows that the operator~\eqref{opo} realizes an isomorphism. In
particular, the estimate
\begin{equation}\label{op}
\|u\|_{ H^2(\Bbb R\times\Omega)}\leq
C(\|(\Delta_\Omega-(1+\lambda)^{-2}\partial_x^2-\mu)u\|_{H^0(\Bbb
R\times\Omega)}+\|\partial_\eta u\|_{H^{1/2}(\Bbb
R\times\partial\Omega)})
\end{equation}
is valid with an independent of $u\in H^2(\Bbb R\times\Omega)$ constant $C=C(\mu,\lambda)>0$.

 Let $\chi_T(x)=\chi(x-T)$, where $\chi\in  C^\infty(\Bbb R)$ is a cutoff function, such that
$\chi(x)=1$ for $x\geq -3$ and $\chi(x)=0$ for $x\leq -4$. As a
consequence of the stabilization of $^\lambda\!\Delta$ and
$^\lambda\!\partial_\nu$ at infinity, we obtain the estimate
$$
\begin{aligned}
\|({^\lambda\!\Delta}-\Delta_\Omega+(1 &
+\lambda)^{-2}\partial_x^2)\chi_T u\|_{ H^0(\Bbb
R\times\Omega)}\\&+\|({^\lambda\!\partial_\nu-\partial_\eta})\chi_T
u\|_{ H^{1/2}(\Bbb R\times\partial\Omega)}\leq c(T)\|\chi_T
u\|_{H^2(\Bbb R\times\Omega)},
\end{aligned}
$$
 where $c(T)\to 0$ as $T\to+\infty$.   This together with~\eqref{op} implies that for a sufficiently large fixed $T=T(\mu,\lambda)>0$ the estimate
\begin{equation}\label{einf}
 \|\chi_Tu\|_{ H^{2}(\Bbb R\times\Omega)} \leq \texttt{C} \bigl(\|(^\lambda\!\Delta -  \mu) \chi_T u\|_{ H^0(\Bbb R\times\Omega)}
 +\|{^\lambda\!\partial_\nu} \chi_T u\|_{ H^{1/2}(\Bbb R\times\partial\Omega)}
\bigr)
\end{equation}
  holds, where the constant $\texttt{C}=(1/C- c(T))^{-1}>0$ may depend on $\mu$ and $\lambda$, but not on $u\in H^2(\Bbb R\times\Omega)$.

Without loss of generality we can  assume that
$(0,T)\times\Omega\subset\mathcal M_c$, cf.~Fig~\ref{manif}. If it
is not the case, then we take a larger smooth compact manifold
$\mathcal M_c$, inserting the cylinder $(0,T)\times\Omega$ instead
of the part $(0,1)\times\Omega$ of $\mathcal M_c$; recall that
$(0,1)\times\Omega\subset\mathcal M_c\cap\Pi$ by our assumptions in
Section~\ref{s2}.

Let $\rho,\varsigma\in C_c^\infty(\mathcal M)$ be some cutoff
functions, such that $\rho= 1$ on $\mathcal M\setminus(T-2,\infty)$
and $\rho= 0$ on $(T-1,\infty)\times\Omega$, while
$\varsigma\rho=\rho$ and $\supp\varsigma\subset\mathcal M_c$. We
assume that $u\in C_c^\infty(\mathcal M)$, and
$\varsigma{^\lambda\!\partial_\nu}u$
 is extended from $\partial\mathcal M_c\cap\partial\mathcal M$ to $\partial\mathcal M_c\setminus\partial\mathcal M$ by zero. As the deformed Neumann Laplacian is an elliptic operator on $\mathcal M$, the local coercive estimate
\begin{equation}\label{ecomp}
\|\rho u\|_{H^2(\mathcal M_c)}\leq C(\|\varsigma{^\lambda\!\Delta}
u\|_{L^2(\mathcal M)}+\|\varsigma
{^\lambda\!\partial_\nu}u\|_{H^{1/2}(\partial\mathcal
M_c)}+\|\varsigma u\|_{L^2(\mathcal M)})
\end{equation}
can be obtained in a usual way from local elliptic coercive
estimates in $\Bbb R^{n+1}$ and $\Bbb R^{n}\times\Bbb R_+$ by gluing
them together with the help of local coordinates and a finite
partition of unity on $(\mathcal M_c,\mathsf
g\!\upharpoonright_{\mathcal M_c})$. (In~\eqref{ecomp} and further
in this proof we write $\|\cdot\|_{L^2(\mathcal M)}$ for the norm in
$L^2(\mathcal M)$ for  uniformity of notations.) We rewrite the
estimate~\eqref{einf} for $u\in C_c^\infty(\mathcal M)$ in the form
$$
\begin{aligned}
\|\chi_Tu\|_{ H^{2}(\Bbb R\times\Omega)} \leq & \texttt{C}
\bigl(\|\chi_T(^\lambda\!\Delta -  \mu)u\|_{H^0(\Bbb
R\times\Omega)}+\|\chi_T{^\lambda\!\partial_\nu}u\|_{H^{1/2}(\Bbb
R\times\partial\Omega)}
\\
&+\|[{^\lambda\!\Delta},\chi_T]u\|_{ H^0(\Bbb R\times\Omega)}
 +\|[{^\lambda\!\partial_\nu}, \chi_T] u\|_{ H^{1/2}(\Bbb R\times\partial\Omega)}
\bigr),
\end{aligned}
$$
where the commutators $[{^\lambda\!\Delta},\chi_T]$ and
$[{^\lambda\!\partial_\nu}, \chi_T]$ are  equal to zero outside of
the set $(T-5,T-2)\times\Omega\subset\mathcal M_c$. Since $\rho=1$
on this set, we get the estimate
$$
\|[{^\lambda\!\Delta},\chi_T]u\|_{ H^0(\Bbb
R\times\Omega)}+\|[{^\lambda\!\partial_\nu}, \chi_T] u\|_{
H^{1/2}(\Bbb R\times\partial\Omega)}\leq C \|\rho u\|_{H^2(\mathcal
M_c)}.
$$
Due to stabilization of $\mathsf g$ at infinity to the product
metric  $dx\otimes dx+\mathfrak h$  we  have
$$
\|\chi_T F\|_{H^0(\Bbb R\times\Omega)}^2=\int_{\Bbb R_+}\|\chi_T
F\|^2_{L^2(\Omega)}\,dx\leq C\|F\|^2_{L^2(\mathcal M)}\quad \forall
F\in L^2(\mathcal M).
$$

Before  proceeding  further, introduce the Sobolev space
$H^2(\mathcal M)$ as the completion of the set $C_c^\infty(\mathcal
M)$ with respect to the norm
$$
\| u\|_{H^2(\mathcal M)}:=\|\chi_Tu\|_{ H^{2}(\Bbb
R\times\Omega)}+\|\rho u\|_{H^2(\mathcal M_c)}.
$$
We also introduce the space of traces $H^{1/2}(\partial\mathcal M)$
as the completion of the set $C_c^\infty(\partial\mathcal M)$ with
respect to the norm
$$
\|u\|_{H^{1/2}(\partial\mathcal M)}=\|\chi_T u\|_{H^{1/2}(\Bbb
R\times\partial\Omega)}+\|\varsigma u\|_{H^{1/2}(\partial\mathcal
M_c)}.
$$
Then from the last four estimates it follows that
\begin{equation}\label{coercive!}
\| u\|_{H^2(\mathcal M)}\leq C\bigl(\|({^\lambda\!\Delta} -
\mu)u\|_{L^2(\mathcal
M)}+\|{^\lambda\!\partial_\nu}u\|_{H^{1/2}(\partial\mathcal
M)}+\|\varsigma u\|_{L^2(\mathcal M)}\bigr),
\end{equation}
where the constant $C$ depends on $\lambda$ and $\mu$, but not on
$u\in H^2(\mathcal M)$.
 Due to the stabilization of $\mathsf g$ at infinity to $dx\otimes dx+\mathfrak h$ we also have the estimate
$$
 \|{^\lambda\!\Delta}u\|_{L^2(\mathcal M)}+ \|{^\lambda\!\partial_\nu}u\|_{H^{1/2}(\partial\mathcal M)}+\|u\|_{L^2(\mathcal M)}\leq c \| u\|_{H^2(\mathcal M)}\quad\forall u\in H^2(\mathcal M),
$$
which can be easily verified in local coordinates. This together
with the estimate~\eqref{coercive!} implies that
$\|\cdot\|_{H^2(\mathcal M)}$ is an equivalent norm in the space
$\mathbf D({^\lambda\!\Delta})$.

Let $\mathsf w$ be a bounded rapidly decreasing at infinity positive
function on $\mathcal M$, such that the embedding of $H^2(\mathcal
M)$ into the weighted space $L^2(\mathcal M,\mathsf w)$ with the
norm $\|\mathsf w \cdot\|_{L^2(\mathcal M)}$ is compact. Then
$\mathbf D({^\lambda\!\Delta})$ is compactly embedded into
$L^2(\mathcal M,\mathsf w)$. As a consequence of~\eqref{coercive!}
we obtain the estimate
\begin{equation}\label{coer}
\| u\|_{H^2(\mathcal M)}\leq {\mathrm C}\bigl(\|({^\lambda\!\Delta}
-  \mu)u\|_{L^2(\mathcal M)}+\|\mathsf w u\|_{L^2(\mathcal
M)}\bigr)\quad\forall u\in\mathbf D({^\lambda\!\Delta})
\end{equation}
of type~\eqref{coercive}. Then by the Peetre's lemma the range of
the continuous operator ${^\lambda\!\Delta} -  \mu:\mathbf
D({^\lambda\!\Delta})\to L^2(\mathcal M)$ is closed and the kernel
is finite-dimensional.

In order to see that the cokernel of the operator
${^\lambda\!\Delta} -  \mu$ is finite dimensional, one can apply the
same argument to the  adjoint operator $^\lambda\!\Delta^*=\frac {1}
{\varrho_{\overline{\lambda}}}{^{\overline{\lambda}}\!\Delta}{\varrho_{\overline{\lambda}}}$,
defined on the functions $u\in H^2(\mathcal M)$ satisfying the
boundary condition $^{\overline{\lambda}}\partial_\nu
\varrho_{\overline{\lambda}} u=0$. We only note that the operators
$\frac {1}
{\varrho_{\overline{\lambda}}}{^{\overline{\lambda}}\!\Delta}{\varrho_{\overline{\lambda}}}$
and $\frac {1}
{\varrho_{\overline{\lambda}}}{^{\overline{\lambda}}\partial_\nu}
\varrho_{\overline{\lambda}}$ stabilize at infinity to the operators
$\Delta_\Omega-(1+\overline{\lambda})^{-2}\partial_x^2$ and
$\partial_\eta$. If $\mu$ does not meet the condition~\eqref{eq9},
this allows to obtain the estimate
$$
\| u\|_{H^2(\mathcal M)}\leq \mathrm
C\bigl(\|({^\lambda\!\Delta}-\mu)^*u\|_{L^2(\mathcal M)}+\|\mathsf w
u\|_{L^2(\mathcal M)}\bigr)\quad\forall u\in\mathbf
D({^\lambda\!\Delta}^*),
$$
 which implies that the cokernel  $\operatorname{coker}({^\lambda\!\Delta} -  \mu)=\ker (^\lambda\!\Delta-\mu)^*$ is finite dimensional. Thus the deformed Neumann Laplacian ${^\lambda\!\Delta}-\mu:\mathbf D({^\lambda\!\Delta})\to L^2(\mathcal M)$ is Fredholm, if $\mu$ does not meet the condition~\eqref{eq9}.

{\it Necessity.} Now we assume that $\mu$ meets the condition~\eqref{eq9} for some $j$, and show that 
 the operator ${^\lambda\!\Delta} -  \mu:\mathbf D({^\lambda\!\Delta})\to L^2(\mathcal M)$ is not Fredholm.  By the Peetre's lemma it suffices to find a sequence $\{v_\ell\}_{\ell=1}^\infty$ of functions $v_\ell\in\mathbf D({^\lambda\!\Delta})$ violating the estimate~\eqref{coer}.

We first show that for some $\mu_0<0$ the continuous operator
\begin{equation}\label{ad}
 \{{^\lambda\!\Delta} -  \mu_0,{^\lambda\!\partial_\nu}\}: H^2(\mathcal M)\to L^2(\mathcal M)\times H^{1/2}(\partial\mathcal M)
\end{equation}
 realizes an isomorphism. With this aim in mind we  replace $\varsigma u$ by $\mathsf w u$ in the estimate~\eqref{coercive!}. Then  by the Peetre's lemma  the range of the operator~\eqref{ad} is closed. It is easy to see that the elements in the cokernel of the operator~\eqref{ad} are of the form $\{v, v\!\upharpoonright_{\partial\mathcal M}\}$, where $v\in \ker({^\lambda\!\Delta} -  \mu_0)^*\subset \mathbf D({^\lambda\!\Delta^*})$. Indeed, let $\{v,\underline v\}$ be in the kernel of the adjoint operator
$$ \{{^\lambda\!\Delta} -  \mu_0,{^\lambda\!\partial_\nu}\}^*: L^2(\mathcal M)\times (H^{1/2}(\partial\mathcal M))^* \to (H^{2}(\mathcal M))^*.
$$
Then $({^\lambda\!\Delta}u-\mu_0 u,v)+\langle
{^\lambda\!\partial_\nu} u, \underline v\rangle=0$ for all $u\in
H^2(\mathcal M)$, where the brackets $\langle\cdot,\cdot\rangle$
denote the inner product in the space $L^2(\partial\mathcal M)$ on
$(\partial\mathcal M,\mathsf g\!\upharpoonright_{\partial\mathcal
M})$ extended to the pairs in $H^{1/2}(\partial\mathcal M)\times
(H^{1/2}(\partial\mathcal M))^*$. Since $\mathbf
D({^\lambda\!\Delta})\subset H^2(\mathcal M)$, we immediately see
that $v\in \ker({^\lambda\!\Delta} -  \mu_0)^*$. Then the Green
identity gives
$$
\langle {^\lambda\!\partial_\nu} u, \underline v -
v\!\upharpoonright_{\partial\mathcal M}\rangle=0\quad \forall u\in
H^2(\mathcal M),
$$
and therefore $\underline v = v\!\upharpoonright_{\partial\mathcal
M}$. By Theorem~\ref{AF}.2  there exists $\mu_0<0$, such that
$\mu_0\notin\sigma({^\lambda\!\Delta})$. As a consequence, the
operator~\eqref{ad} realizes an isomorphism.

Let $\chi$ be a smooth cutoff function on the real line, such that
$\chi(x)=1$ for $|x-3|\leq 1$, and $\chi(x)=0$ for $|x-3|\geq 2$.
Consider the functions
\begin{equation}\label{test}
u_\ell(x,\mathrm
y)=\chi(x/\ell)\exp\bigl({i(1+\lambda)\sqrt{\mu-\nu_j}x}\bigr)\Phi(\mathrm
y),\quad (x,\mathrm y)\in \Bbb R\times\Omega,
\end{equation}
where $\Phi$ is an eigenfunction of the Neumann Laplacian
$\Delta_\Omega$ corresponding to the eigenvalue $\nu_j$. It is clear
that $u_\ell$ satisfies the Neumann boundary condition
$\partial_\eta u_\ell=0$ on $\Bbb R\times\partial\Omega$. As $\mu$
meets the condition~\eqref{eq9}, the exponent in~\eqref{test} is an
oscillating function of $x\in\Bbb R$. Straightforward calculation
shows that
\begin{equation}\label{s--}
\bigl\|\bigl(\Delta_\Omega-(1+  \lambda)^{-2}\partial_x^2 -\mu\bigr)
u_\ell\bigr\|_ {H^0(\Bbb R\times\Omega)}\leq const,\quad\|u_\ell\|_{
H^2(\Bbb R\times\Omega)}\to\infty
\end{equation}
as $\ell\to +\infty$. We extend the functions $u_\ell$ from their
supports in $\Pi$ to $\mathcal M$ by zero. Then $u_\ell\in
C_c^\infty(\mathcal M)$ and $^\lambda\!\partial_\nu u_\ell\in
C_c^\infty(\partial\mathcal M)$.

Introduce the functions
$$
v_\ell=u_\ell-(\{{^\lambda\!\Delta} -
\mu_0,{^\lambda\!\partial_\nu}\})^{-1}\{0,^\lambda\!\partial_\nu
u_\ell\}.
$$
It is clear that $v_\ell\in\mathbf D({^\lambda\!\Delta})$. We also
have
\begin{equation}\label{s-}
\begin{aligned}
\|(\{{^\lambda\!\Delta} -  &
\mu_0,{^\lambda\!\partial_\nu}\})^{-1}\{0,^\lambda\!\partial_\nu
u_\ell\}\|_{H^2(\mathcal M)}
\\
& \leq C
\|({^\lambda\!\partial_\nu}-\partial_\eta)u_\ell\|_{H^{1/2}(\partial\mathcal
M)}\leq \texttt{C}_\ell\|u_\ell\|_{ H^2(\Bbb R\times\Omega)},
\end{aligned}
\end{equation}
where $\texttt{C}_\ell\to 0$ as $\ell\to+\infty$ due to
stabilization of $^\lambda\!\partial_\nu$ to $\partial_\eta$ at
infinity. Hence
\begin{equation}\label{s+}
\begin{aligned}
&\|v_\ell\|_{H^2(\mathcal M)}\geq \|u_\ell\|_{ H^2(\Bbb
R\times\Omega)}-\texttt{C}_\ell\|u_\ell\|_{ H^2(\Bbb
R\times\Omega)},
\\
&\|v_\ell-u_\ell\|_{H^2(\mathcal M)}\leq \texttt{C}_\ell\|u_\ell\|_{
H^2(\Bbb R\times\Omega)}.
\end{aligned}
\end{equation}
Assume that the estimate~\eqref{coer} is valid. Without loss of
generality we can take a rapidly decreasing weight $\mathsf w$, such
that $\|\mathsf w u_\ell\|_{L^2(\mathcal M)}\leq Const$ uniformly in
$\ell\geq 1$, and the imbedding $H^2(\mathcal M) \hookrightarrow
L^2(\mathcal M;\mathsf w)$ is compact. It is clear that $\|\mathsf w
u\|_{L^2(\mathcal M)}\leq c\|u\|_{H^2(\mathcal M)}$ with some
independent of $u\in H^2(\mathcal M)$ constant $c$. Therefore
\begin{equation}\label{s.}
\|\mathsf w v_\ell\|_{L^2(\mathcal M)}\leq Const + c
\texttt{C}_\ell\|u_\ell\|_{ H^2(\Bbb R\times\Omega)},
\end{equation}
cf.~\eqref{s+}. Due to stabilization of  ${^\lambda\!\Delta}$ to
$\Delta_\Omega-(1+  \lambda)^{-2}\partial_x^2$ at infinity we have
$$
\|({^\lambda\!\Delta}-\Delta_\Omega+(1+
\lambda)^{-2}\partial_x^2)u_\ell\|_{L^2(\mathcal
M)}\leq\texttt{c}_\ell\|u_\ell\|_{H^2(\Bbb R\times\Omega)},
$$
 where $\texttt{c}_\ell\to 0$ as $\ell\to+\infty$. This together with~\eqref{s--} and~\eqref{s+}  gives
\begin{equation}\label{s++}
\begin{aligned}
\|({^\lambda\!\Delta}-\mu)v_\ell\|_{L^2(\mathcal M)}\leq
C\bigl\|\bigl(-(1+  \lambda)^{-2}\partial_x^2+\Delta_\Omega
-\mu\bigr) u_\ell\bigr\|_ {H^0(\Bbb R\times\Omega)}\\
+\|({^\lambda\!\Delta}-\Delta_\Omega+(1+
\lambda)^{-2}\partial_x^2)u_\ell\|_{L^2(\mathcal
M)}+\|({^\lambda\!\Delta}-\mu)(v_\ell-u_\ell)\|_{L^2(\mathcal M)}
\\
\leq C\cdot const+(\texttt{c}_\ell+\mathsf{C}
\texttt{C}_\ell)\|u_\ell\|_{H^2(\Bbb R\times\Omega)}.
\end{aligned}
\end{equation}
Finally, as a consequence of~\eqref{s+},~\eqref{coer},~\eqref{s++}
and~\eqref{s.},  we get
\begin{equation}\label{final}
\begin{aligned}
\|u_\ell\|_{ H^2(\Bbb R\times\Omega)}-\texttt{C}_\ell\|u_\ell\|_{
H^2(\Bbb R\times\Omega)}
\\
\leq\|v_\ell\|_{H^2(\mathcal M)}\leq \mathrm
C(\|({^\lambda\!\Delta}-\mu)v_\ell\|_{L^2(\mathcal M)}+\|\mathsf w
v_\ell\|_{L^2(\mathcal M)})
\\
\leq \mathrm C( C\cdot const +(\mathsf
C\texttt{C}_\ell+\texttt{c}_\ell)\|u_\ell\|_{ H^2(\Bbb
R\times\Omega)}+Const + c \texttt{C}_\ell\|u_\ell\|_{ H^2(\Bbb
R\times\Omega)}).
\end{aligned}
\end{equation}
Since $\texttt{c}_\ell\to 0$ and $\texttt{C}_\ell\to 0$, the
inequalities~\eqref{final} imply that the value $\|u_\ell\|_{
H^2(\Bbb R\times\Omega)}$ remains bounded as $\ell\to+\infty$. This
contradicts~\eqref{s--}. Thus the sequence
$\{v_\ell\}_{\ell=1}^\infty$ violates the estimate~\eqref{coer}. The
necessity is proven. \qed\end{pf*}

\section{Resolvent matrix elements meromorphic continuation}\label{s6}

In this section we complete the proof of Theorem~\ref{T1}, applying
the Aguilar-Balslev-Combes argument to the resolvent matrix elements
$((\Delta-\mu)^{-1}F,G)$, where $F$  and $G$ are in the set
$\mathcal A$ of analytic vectors, and $(\cdot,\cdot)$ is the inner
product in $L^2(\mathcal M)$.

Recall that $\mathscr E$ is the algebra of all entire functions
$\mathbb C\ni z\mapsto f(z,\cdot)\in C^\infty(\Omega)$, such that in
any sector $|\Im z|\leq (1-\epsilon) \Re z$ with $\epsilon>0$  the
value $\|f(z,\cdot)\|_{L^2(\Omega)}$ decays faster than any  inverse
power of $\Re z$  as $\Re z\to+\infty$. By definition a function
$F\in L^2(\mathcal M)$ is in the set $\mathcal A$, if $F(x,\mathrm
y)=f(x,\mathrm y)$  for some $f\in\mathscr E$ and all $(x,\mathrm
y)\in\Pi$. For $F\in\mathcal A$ and $\lambda\in\mathbb C$  we define
the function  $F\circ\varkappa_\lambda$ on $\mathcal M$ by the rule
$F\circ\varkappa_\lambda\equiv F$ on $\mathcal M\setminus\Pi$, and
\begin{equation}\label{eq}
F\circ\varkappa_\lambda(x,\mathrm y)=f(x+\lambda \mathsf
s_R(x),\mathrm y),\quad (x,\mathrm y)\in\Pi.
\end{equation}
Here $f(x+\lambda \mathsf s_R(x),\cdot)$ is the value of the
corresponding to $F$ entire function $f\in\mathscr E$ at the point
$z=x+\lambda \mathsf s_R(x)$, and $\varkappa_\lambda(x,\mathrm
y)=(x+\lambda \mathsf s_R(x),\mathrm y)$ is the complex scaling in
$\Pi$.

\begin{lem}\label{p1} Let $\mathsf s_R(x)=\mathsf s(x-R)$, where $R>0$ and $\mathsf s$ a smooth function   obeying the conditions~\eqref{ab}. Then we have:
{\rm 1)} For any $F\in\mathcal A$, $\lambda\mapsto
F\circ\varkappa_\lambda$ is an $L^2(\mathcal M)$-valued analytic
function in the disk $|\lambda|<2^{-1/2}$; {\rm 2)} For any
$\lambda$ in this disk the image $\varkappa_\lambda[\mathcal
A]=\{F\circ\varkappa_\lambda: F\in\mathcal A\}$ of $\mathcal A$
under $\varkappa_\lambda$ is dense in the space $L^2(\mathcal M)$.
\end{lem}
\begin{pf}  In essence, this proposition is based on~\cite[Theorem 3]{Hunziker}.

Since $F\circ\varkappa_\lambda\equiv F$ on $\mathcal M\setminus\Pi$,
it suffices to consider the functions $F\in\mathcal A$ with supports
in the semi-cylinder $\Pi$. For all these functions the estimate
\begin{equation}\label{re}
\|F\|^2\leq  C \int_{\Bbb R_+}\|F(x,\cdot)\|_{L^2(\Omega)}^2\,dx
\end{equation}
is valid due to the stabilization of $\mathsf g$ to the product
metric $dx\otimes dx+\mathfrak h$ at infinity; recall that
$\|\cdot\|$ is  the norm in the space $L^2(\mathcal M)$ on the
Riemannian manifold $(\mathcal M,\mathsf g)$.

{\rm {1)}} Let $x\in\Bbb R_+$.  We set $z=x+\lambda \mathsf s_R(x)$.
Then $|\Re z|^2-|\Im z|^2\geq x^2/2-|\lambda|^2|\mathsf s_R(x)|^2$.
Since $\mathsf s_R(x)<x$, for all $\lambda$ in the disk
$|\lambda|\leq\sqrt{ 1/2-\epsilon}$ we get
$$
|\Re z|^2-|\Im z|^2\geq \epsilon x^2\geq \epsilon |\mathsf
s_R(x)|^2\geq 2\epsilon |\Im z|^2.
$$
Therefore $|\Im z|\leq (1+2\epsilon)^{-1/2}\Re z$. On the other hand
$\Re z\geq (1-2^{-1/2})x$. The equality~\eqref{eq} with
$f\in\mathscr E$, combined with the definition of the
algebra~$\mathscr E$, implies that the value
$\|F\circ\varkappa_\lambda(x,\cdot)\|_{ L^2(\Omega)}$ decreases
faster than any inverse power of $x$ as $x\to\infty$, uniformly in
$\lambda$ with $|\lambda|\leq\sqrt{1/2-\epsilon}$. Therefore we have
$F\circ\varkappa_\lambda\in L^2(\mathcal M)$ because of~\eqref{re}.
 It remains to note that $( F\circ\varkappa_\lambda,G)$ is analytic in $|\lambda|<2^{-1/2}$ for any
$ G \in L^2(\mathcal M )$, where $(\cdot,\cdot)$ is the inner
product in $L^2(\mathcal M)$. The assertion 1) of the proposition is
proven.

 {\rm 2)} Given $h\in C^\infty_c(\Pi)$ we  construct
 a sequence $f_\ell\in\mathscr E$, such that  the
 function $\mathbb R_+\ni x\mapsto g_\ell(x)=f_\ell(x+\lambda \mathsf s_R(x))\in C^\infty(\Omega)$ tends to $h$ in
 $L^2({\mathcal M})$ as $\ell\to \infty$. Since the set $\{F\in L^2(\mathcal M): F\!\upharpoonright_{\Pi}\in C_c^\infty(\Pi)\}$ is dense in $L^2(\mathcal M)$, this will imply that the set $\varkappa_\lambda[\mathcal A]$ is also dense in $L^2(\mathcal M)$.

 Namely, let
$$
f_\ell(z)=\sqrt{\frac {\ell} {\pi}}\int_{\mathbb R}  h(x)\exp[-\ell
(z-x-\lambda \mathsf s_R(x))^2](1+\lambda \mathsf s_R'(x))\,dx,\
\ell\geq 1,
$$
where $h(x)=0$ for $x\leq 0$. It is clear that $z\mapsto
 f_\ell(z)\in C^\infty(\Omega)$ is an entire function. Since $h$ is compactly supported, $z\mapsto \|f_\ell(z)\|_{ L^2(\Omega)}$  has the same falloff at infinity as $\exp(-\ell z^2)$, i.e. $f_\ell\in\mathscr E$.
In order to prove that $f_\ell$ tends to $h$ in $L^2(\mathcal M)$,
we set
$$
v(x,\tilde x;\lambda)=x+\lambda \mathsf s_R(x)-\tilde x-\lambda
\mathsf s_R(\tilde x).
$$
From the conditions~\eqref{ab} on the function $\mathsf s$ it
follows that for all $\lambda$ in the disk $|\lambda|\leq
\sqrt{1/2-\epsilon}$ we get $|\Re v|^2-|\Im v|^2\geq
\epsilon|x-\tilde x|^2$, and therefore
\begin{equation}\label{star}
|\exp(-v^2(x,\tilde x;\lambda))|\leq\exp(-\epsilon|x-\tilde x|^2).
\end{equation}
For all real $\lambda$ in the disk $|\lambda|\leq\sqrt{
1/2-\epsilon}$ we get the equalities
$$
\sqrt{\frac {\ell} {\pi}}\int_{\mathbb R} \exp[-\ell (x+\lambda
\mathsf s_R(x)-\tilde x-\lambda \mathsf s_R(\tilde x))^2](1+\lambda
\mathsf s_R'(x))\,dx=\sqrt{\frac {\ell} {\pi}}\int_{\mathbb
R}e^{-\ell v^2}\,dv=1.
$$
Due to~\eqref{star}  these equalities extend by analyticity to the
disk $|\lambda|\leq \sqrt{1/2-\epsilon}$. Thus we have established
the equality
\begin{equation}\label{eq14}
h(\tilde x)-g_\ell(\tilde x)=\sqrt{\frac {\ell} {\pi}}\int_{\mathbb
R}e^{-\ell v^2(x,\tilde x;\lambda)}(h(\tilde x)-h(x))(1+\lambda
\mathsf s_R'(x))\, dx.
\end{equation}
This together with~\eqref{star} gives us the uniform in $\lambda$
estimate
\begin{equation}\label{eq13}
\|h(\tilde x)-g_\ell(\tilde x)\|_{ L^2(\Omega)}\leq C
\sqrt{\ell}\int_{\mathbb R}e^{-\ell\varepsilon (x-\tilde x)^2}
\|h(\tilde x)-h(x)\|_{ L^2(\Omega)}\,dx.
\end{equation}
It is known property of the Weierstra{\ss} singular
integral~\cite{Titchmarsh} that for all $\tilde x\in\mathbb R$
\begin{equation}\label{eq13+}
\sqrt{\ell}\int_{\mathbb R}e^{-\ell\varepsilon (x-\tilde
x)^2}\|h(\tilde x)-h(x)\|_{ L^2(\Omega)} \, dx \to 0\ \text{ as }\
\ell\to+\infty.
\end{equation}
 From~\eqref{eq13},~\eqref{eq13+}, and~\eqref{re} we conclude that $g_\ell$
 converges to $h$ in the norm of $L\sp 2({\mathcal M})$ as $\ell\to+\infty$.
\qed\end{pf}

\begin{pf*}{\bf Proof of Theorem~\ref{T1}.} The assertion~{\it 1} was proven in Theorem~\ref{AF},  and the assertion~{\it 3} is a direct consequence of Theorem~\ref{ess}.

{\it 4.} Let $\lambda\in\mathcal D_\alpha$ be fixed.  By
Theorem~\ref{AF}.2  there is a point $\mu<0$ in the resolvent set of
$^\lambda\!\Delta$. Every $\mu<0$ is in the simply connected set
$\mathbb C\setminus\sigma_{ess}({^\lambda\!\Delta})$, cf.
Fig.~\ref{fig5}.  This  implies that ${^\lambda\!\Delta}-\mu:\mathbf
D( {^\lambda\!\Delta})\to L^2(\mathcal M)$ is an analytic  Fredholm
operator function of $\mu\in\mathbb
C\setminus\sigma_{ess}({^\lambda\!\Delta})$. It is a known result of
the analytic Fredholm theory e.g.~\cite[Proposition
A.8.4]{KozlovMaz`ya} that the spectrum  of an analytic Fredholm
operator function consists of isolated eigenvalues of finite
algebraic multiplicity. Thus $\sigma
({^\lambda\!\Delta})=\sigma_{ess}({^\lambda\!\Delta})\cup \sigma_d
({^\lambda\!\Delta})$.

{\it 6.} Let us first obtain a relation between the matrix elements
$((\Delta-\mu)^{-1}F,G)$ and some matrix elements of the resolvent
$({^\lambda}\Delta-\mu)^{-1}$ for a real $\lambda\in\mathcal
D_\alpha$.

 The Riemannian geometry gives the equality
\begin{equation}\label{asd}
(\Delta
-\mu)u=\bigl(({^\lambda\!\Delta}-\mu)(u\circ\varkappa_\lambda)\bigr)\circ\varkappa_\lambda^{-1}\quad
\forall u\in \mathbf C({^0\!\Delta}),
\end{equation}
 where $u\circ\varkappa_\lambda$ is in the core $\mathbf C({^\lambda\!\Delta})$ introduced in Definition~\ref{d1}. We take  $\mu<0$ outside of the sector of the m-sectorial operator ${^\lambda\!\Delta}$, see  Theorem~\ref{AF}.2. Then for all $\lambda\in\mathcal D_\alpha$ the operator ${^\lambda\!\Delta}-\mu$ has a bounded inverse, and we can rewrite the equality~\eqref{asd} in the form
\begin{equation}\label{++}
(\Delta
-\mu)^{-1}F=\bigl(({^\lambda\!\Delta}-\mu)^{-1}(F\circ\varkappa_\lambda)\bigr)\circ\varkappa_\lambda^{-1},
\end{equation}
where $F$ is in the set $\{F=(\Delta-\mu)u: u\in \mathbf
C({^0\!\Delta})\}$. This set is dense in $L^2(\mathcal M)$, because
the core $\mathbf C({^0\!\Delta})$ is dense in $L^2(\mathcal M)$,
and the operator $\Delta-\mu$ is invertible. Recall that $
(F,G)_\lambda=\int F\overline{G}\,\operatorname{dvol}_{\lambda}$ for
$F,G\in L^2(\mathcal M)$, and $\sqrt{(\cdot,\cdot)_\lambda}$  is an
equivalent norm in $L^2(\mathcal M)$ as $\lambda$ is real,
see~Section~\ref{s3}. It is clear that
$(F\circ\varkappa_\lambda,F\circ\varkappa_\lambda)_\lambda=(F,F)_0$.
As a consequence, the (real) scaling $F\mapsto
F\circ\varkappa_\lambda$ realizes an isomorphism in $L^2(\mathcal
M)$, and the equality~\eqref{++} extends by continuity to all $F\in
L^2(\mathcal M)$. Taking the inner product in $L^2(\mathcal M)$ of
the equality~\eqref{++} with $G\in L^2(\mathcal M)$, and using the
identity $(H\circ\varkappa^{-1}_\lambda,
G)=(H,G\circ\varkappa_\lambda)_\lambda$ in the right hand side,  we
obtain the relation
\begin{equation}\label{h6}
\bigl((\Delta-\mu)^{-1}F,G
\bigr)=\bigl(({^\lambda\!\Delta}-\mu)^{-1}(F\circ\varkappa_\lambda),G\circ\varkappa_{\overline{\lambda}}
\bigr)_\lambda, \quad\lambda\in \mathcal D_\alpha\cap\mathbb R,
\end{equation}
between the matrix elements of the resolvents $(\Delta-\mu)^{-1}$
and $({^\lambda\!\Delta}-\mu)^{-1}$.

We intend to implement the Aguilar-Balslev-Combes argument to the
equality~\eqref{h6}. In other words, for arbitrary $F$ and $G$  in
the set $\mathcal A$ of analytic vectors, we will extend the
equality~\eqref{h6} by analyticity to all $\lambda$ in the disk
$\mathcal D_\alpha$. Then the right hand side of~\eqref{h6} will
provide the left hand side with a meromorphic continuation in $\mu$
across  $\sigma_{ess}({\Delta})$.

Let $F,G\in\mathcal A$. Then by Lemma~\ref{p1}
$F\circ\varkappa_\lambda$ and $G\circ\varkappa_\lambda$ are
$L^2(\mathcal M)$-valued analytic functions of $\lambda$ in the disk
$\mathcal D_\alpha$.  The resolvent $({^\lambda\!\Delta}-\mu)^{-1}$
is an analytic function of $\lambda\in\mathcal D_\alpha$ by
Theorem~\ref{AF}.3.  As a consequence, the equality~\eqref{h6}
extends by analyticity to all $\lambda\in\mathcal D_\alpha$.

We take some $\lambda\in\mathcal D_\alpha$ with $\Im\lambda\neq 0$,
and consider the Left Hand Side (LHS) and the Right Hand Side (RHS)
in~\eqref{h6} as functions of $\mu$.   The LHS is meromorphic in
$\mu\in\mathbb C\setminus\sigma_{ess}(\Delta)$ with poles at the
points of $\sigma_{d}(\Delta)$. The spectrum  $\sigma(\Delta)$ of
the nonnegative selfadjoint operator  $\Delta$ is a subset of the
half-line $\overline{\mathbb R}_+$. On the other hand,  the RHS is a
meromorphic function  on the set $\mu\in\mathbb
C\setminus\sigma_{ess}({^\lambda\!\Delta})$. Therefore the RHS
provides the LHS with a meromorphic continuation from $\mu\in\mathbb
C\setminus\sigma_{ess}(\Delta)$ across the intervals
$(\nu_j,\nu_{j+1})$, $j\in\mathbb N$, to the strips between the rays
of $\sigma_{ess}({^\lambda\!\Delta})$, cf. Fig.~\ref{fig5}. It is
clear that the meromorphic continuation  can have poles only  at
points  of $\sigma_d({^\lambda\!\Delta})$. Conversely, let
$\mu_0\in\sigma_d({^\lambda\!\Delta})$, and let $\mathsf P$ be the
corresponding Riesz projection (i.e. the first order residue of
$({^\lambda\!\Delta}-\mu)^{-1}$ at the pole $\mu_0$). The kernel
$\ker({^\lambda\!\Delta}-\mu_0)\neq\{0\}$ is in the range of
$\mathsf P$.  Recall that the form $(\cdot,\cdot)_\lambda$  is
non-degenerate, and the sets $\varkappa_\lambda[\mathcal A]$ and
$\varkappa_{\overline{\lambda}}[\mathcal A]$ are dense in
$L^2(\mathcal M)$ by Lemma~\ref{p1}. Therefore for some $F,
G\in{\mathcal A}$ we must have $ \bigl(\mathsf P
F\circ\varkappa_\lambda,G\circ\varkappa_{\overline{\lambda}}\bigr)_\lambda\neq
0$. Thus $\mu_0$ is a pole.

{\it 2.} The LHS is independent of the scaling function $\mathsf s$.
Hence the meromorphic continuation of LHS and its poles are
independent of $\mathsf s$. This together with the assertion~{\it 6}
implies that $\sigma_d({^\lambda\!\Delta})$ does not depend on
$\mathsf s$. By the assertion~{\it 3} the essential spectrum
$\sigma_{ess}({^\lambda\!\Delta})$ is also independent of $\mathsf
s$. Therefore the spectrum $\sigma
({^\lambda\!\Delta})=\sigma_{ess}({^\lambda\!\Delta})\cup \sigma_d
({^\lambda\!\Delta})$ is independent of $\mathsf s$.

{\it 5.} Let $\mu\in\sigma_d({^\lambda\!\Delta})$. As $\lambda$
changes continuously in the disk $\mathcal D_\alpha$ and
$\mu\notin\sigma_{ess}({^\lambda\!\Delta})$, the RHS of~\eqref{h6}
provides the LHS with one and the same meromorphic continuation to a
neighborhood of $\mu$. Therefore $\mu$ remains a pole of the
meromorphic continuation, which is equivalent to the inclusion
$\mu\in\sigma_d({^\lambda\!\Delta})$ by the assertion~{\it 6}.

{\it 7.}  Let $\lambda$ be a non-real number in the disk $\mathcal
D_\alpha$. Consider the projection
$$
\mathsf P=\operatorname{s-}\!\lim_{\epsilon\downarrow 0} i\epsilon
(\Delta-\mu_0-i\epsilon)^{-1}
$$
  onto the eigenspace of
the selfadjoint operator $\Delta$. Suppose that $\mu_0\in \mathbb R$
and $\mu_0\notin\sigma({^\lambda \!\Delta})$ (then $\mu_0\neq \nu_j$
for all $j\in\mathbb N$). Therefore, for any $F,G\in\mathcal A$ the
RHS of~\eqref{h6} is an analytic function of $\mu$ in a neighborhood
of $\mu_0$. The equality~\eqref{h6} implies that $(\mathsf P
F,G)=0$. The set $\mathcal A$ is dense in $L^2(\mathcal M)$, and
hence $\mathsf P=0$. Thus $\ker(\Delta-\mu_0)=\{0\}$.

Now we assume that $\mu_0\in\mathbb R$ and
$\mu_0\in\sigma_d({^\lambda \!\Delta})$. Then the resolvent
$({^\lambda \!\Delta}-\mu)^{-1}$ has a pole at $\mu_0$. The sets
$\varkappa_\lambda[\mathcal A]$ and
$\varkappa_{\overline{\lambda}}[\mathcal A]$ are dense in
$L^2(\mathcal M)$. Hence there exist $F,G\in \mathcal A$, such that
$\mu_0$ is a pole for the RHS of~\eqref{h6}. The equality~\eqref{h6}
implies that $(\mathsf P F,G)\neq 0$, and thus
$\ker(\Delta-\mu_0)\neq\{0\}$.

{\it 8.} The RHS of~\eqref{h6} with $\Im\lambda>0$, and therefore
the LHS, being defined on the dense subset $\mathcal A$ of
$L^2(\mathcal M)$,  has limits at the points $\Bbb
R\setminus\sigma({^\lambda\!\Delta})$
 as $\mu$ tends to the real line from $\mathbb C^+$. Since the set $\Bbb R\cap\sigma({^\lambda\!\Delta})$ is countable,  the  Laplacian
$\Delta$ has no singular continuous spectrum,
e.g.~\cite[Theorem~XII.20]{Simon Reed iv}.
 \qed\end{pf*}

\end{document}